\documentclass{JHEP3}

\usepackage{amsmath}
\usepackage{bm}
\usepackage{bbm}
\usepackage{graphicx}
\usepackage{mathrsfs}
\usepackage{slashed}

\newcommand{\threeint}[1]{\int \!\! \frac{\mathrm{d}^3#1}{(2\pi)^3}\,}
\newcommand{\imag}{\mathrm{i}}
\newcommand{\dd}{\mathrm{d}}
\newcommand{\cc}{\mathscr{C}}
\newcommand{\hc}{\text{H.c.}}
\newcommand{\La}{\mathscr{L}}
\newcommand{\Ga}{\mathscr{G}}
\newcommand{\Sa}{\mathscr{S}}
\newcommand{\Za}{\mathscr{Z}}
\newcommand{\Pa}{\mathscr{P}}
\newcommand{\Ta}{\mathscr{T}}
\newcommand{\Da}{\mathscr{D}}
\newcommand{\bs}[1]{\bm{#1}}
\newcommand{\psib}{\bar{\psi}}
\newcommand{\Psib}{\overline{\Psi}}
\newcommand{\psic}{\psi^{\cc}}
\newcommand{\Psic}{\Psi^{\cc}}
\newcommand{\psicb}{\overline{\psi^{\cc}}}
\newcommand{\Psicb}{\overline{\Psi^{\cc}}}
\newcommand{\gr}[1]{\mathop{\rm #1}}
\newcommand{\mfave}[1]{\langle #1\rangle_{\text{mf}}}
\DeclareMathOperator{\tr}{Tr}
\DeclareMathOperator{\re}{Re}
\DeclareMathOperator{\im}{Im}
\newcommand{\openone}{\mathbbm{1}}

\title{QCD-like theories at nonzero temperature and density}

\author{Tian~Zhang,$^a$ Tom\'{a}\v{s} Brauner$^a$\thanks{On leave from
Department of Theoretical Physics, Nuclear Physics Institute ASCR, \v{R}e\v{z},
Czech Republic.}
\space and Dirk H.~Rischke$^{a,b}$\\
${}^a$Institute for Theoretical Physics, Goethe University,
Frankfurt am Main, Germany\\
${}^b$Frankfurt Institute for Advanced Studies,
Frankfurt am Main, Germany\\
E-mail: \email{tzhang@th.physik.uni-frankfurt.de,
brauner@th.physik.uni-frankfurt.de, drischke@th.physik.uni-frankfurt.de}}

\abstract{We investigate the properties of hot and/or dense matter in QCD-like
theories with quarks in a (pseudo)real representation of the gauge group using
the Nambu--Jona-Lasinio model. The gauge dynamics is modeled using a simple
lattice spin model with nearest-neighbor interactions. We first keep our
discussion as general as possible, and only later focus on theories with
adjoint quarks of two or three colors. Calculating the phase diagram in the
plane of temperature and quark chemical potential, it is qualitatively
confirmed that the critical temperature of the chiral phase transition is much
higher than the deconfinement transition temperature. At a chemical potential
equal to half of the diquark mass in the vacuum, a diquark Bose--Einstein
condensation (BEC) phase transition occurs. In the two-color case, a
Ginzburg--Landau expansion is used to study the tetracritical behavior around
the intersection point of the deconfinement and BEC transition lines, which are
both of second order. We obtain a compact expression for the expectation value
of the Polyakov loop in an arbitrary representation of the gauge group (for any
number of colors), which allows us to study Casimir scaling at both nonzero
temperature and chemical potential.}

\preprint{INT-PUB-10-018}

\begin{document}

\section{Introduction}
\label{Sec:intro}

Quantum chromodynamics (QCD) is accepted as the theory for strongly interacting
matter. However, due to the strong coupling, perturbative treatments fail at an
energy scale of the order of $\Lambda_{\text{QCD}}$, resulting in the fact that
the structure of the QCD vacuum is still far from being well understood. The
same is true of the QCD phase diagram where, due to the formidable sign
problem, standard Monte-Carlo techniques based on importance sampling cannot be
used at nonzero quark chemical potential.

In order to get deeper insight into the behavior of dense quark matter, several
QCD-like theories have been proposed, including QCD with an imaginary chemical
potential~\cite{Alford:1998imaginary}, at nonzero isospin
density~\cite{Son:2000qcd}, two-color QCD with fundamental
quarks~\cite{Kogut:1999on}, and any-color QCD with adjoint
quarks~\cite{Kogut:2000qcdlike}. All these theories share the pleasing feature
that they are free of the sign problem, which makes them possible to be
simulated on the lattice from first principles.

In the present paper we focus on two wide classes of QCD-like theories: those
with nonzero \emph{baryon} chemical potential and quarks in a \emph{real} or
\emph{pseudoreal} representation of the gauge group. For the sake of brevity,
they will be henceforth referred to as \emph{type-I and type-II QCD-like
theories}, respectively (see, e.g., Refs.~\cite{Peskin:1980gc,Bijnens:2009qm}
for applications of these theories in another, electroweak, sector of the
standard model of elementary particles). The typical examples are QCD with
adjoint quarks of two (aQC$_2$D) or three (aQCD) colors for type~I, and
two-color QCD with fundamental quarks (QC$_2$D) for type~II.  While the
effective Nambu--Jona-Lasinio (NJL) model description for the latter was worked
out in Refs.~\cite{Kondratyuk:1991hf,Ratti:2004thermodynamics,Sun:2007fc,%
Brauner:2009twocolor,Andersen:2010phase}, the model Lagrangian for type-I
theories is, as far as we know for the first time, constructed here.\footnote{A
Polyakov loop NJL-type model for adjoint fermions was already worked out in
Ref.~\cite{Nishimura:2009me}, but the quark sector of their Lagrangian does not
have the $\gr{SU}(2N_f)$ symmetry of the underlying gauge theory.} Due to the
(pseudo)reality of the quark representation of the gauge group, all these
theories have several remarkable differences as compared to ordinary QCD,
besides the absence of the sign problem itself.

First, with $N_f$ massless quark flavors, the global flavor symmetry is
$\gr{SU}(2N_f)$ rather than the usual chiral group $\gr{SU}(N_f)_{\gr L} \times
\gr{SU}(N_f)_{\gr R} \times \gr{U(1)_B}$. The reason is that the
charge-conjugated quark field $(\psi_{\gr R})^{\cc}$ (charge conjugation being
defined as $\psi^\cc=C\psib^T$) which is a left-handed spinor transforms in the
same way as the left-handed quark $\psi_{\gr L}$ under both color and Lorentz
transformations, so it is allowed to transform them into each other while
keeping the color symmetry intact. This means that the multiplets of states in
the spectrum will contain modes of different baryon number. In particular,
apart from the pions the Nambu--Goldstone (NG) bosons of the spontaneously
broken flavor symmetry will also include diquarks. These light diquarks are
colorless bosons carrying baryon charge, and hence at low temperature and
sufficiently high chemical potential, they will undergo Bose--Einstein
condensation (BEC).

Second, in case of quarks in a real (such as the adjoint) representation, the
$\gr Z_{N_c}$ center symmetry remains intact in the presence of dynamical
quarks. This leads to a well-defined deconfinement phase transition,
accompanied by spontaneous center symmetry breaking, instead of a crossover as
in QCD \cite{Mocsy:2003qw}. The associated order parameter is the expectation
value of the Polyakov loop. For the two- and three-color cases investigated in
this paper, the deconfinement transition is of second and of first order,
respectively.

Since the BEC and deconfinement phase transitions are both well-defined, being
associated with exact symmetries even in the presence of massive dynamical
quarks, aQC$_2$D exhibits a rather unusual critical behavior in the vicinity of
the tetracritical point where the two second-order transition lines cross each
other~\cite{Sannino:2004tetracritical}. In (three-color) aQCD the deconfinement
transition is of first order. As a consequence the second-order BEC critical
line is interrupted around the deconfinement transition, meeting the
deconfinement line at two tricritical points. This general expectation is
confirmed by our explicit model calculation.

To model the gauge sector, we use a simple lattice spin model with
nearest-neighbor interaction, inspired by the strong-coupling
expansion~\cite{Gocksch:1984xc,Dumitru:2003deconfining,Fukushima:2007a,%
Fukushima:2003fm,Gupta:2007ax,Abuki:2009gauge}. This is then coupled to
continuum quarks in a fashion similar to the Polyakov-loop NJL (PNJL)
model~\cite{Abuki:2009gauge,Fukushima:2004chiral,Megias:2004hj,Ratti:2005phases,%
Roessner:2006polyakov,Fukushima:2008phase}. As an effective model, the NJL
model successfully describes chiral symmetry breaking and pairing of quarks.
But since it contains no dynamical gluons, the confinement feature is missing.
In order to account for confinement at least in a heuristic way, one adds to
the thermodynamic potential an effective potential for the Polyakov loop,
adjusted to reproduce the thermodynamic observables in the pure gauge theory.
The Polyakov loop is represented by a constant temporal background gluon field
which in turn couples to the quarks~\cite{Fukushima:2004chiral}. The parameters
of the PNJL model are fixed separately in the pure gauge and chiral quark
sectors. The successful qualitative reproduction of the coincidence of the
deconfinement and chiral restoration temperatures, $T_d$ and $T_\chi$, in QCD
is then one of the great virtues of the PNJL model. On the other hand, aQCD is
very different. First, $T_d \ll T_\chi$, resulting in a broad range of
temperatures exhibiting deconfined, but still chirally broken
matter~\cite{Kogut:1985xa,Karsch:1998deconfinement,Engels:2005te} (see also
Refs.~\cite{Nishimura:2009me,Unsal:2007vu} for related theories with periodic
boundary conditions for quarks). Second, $T_d$ does not change much compared to
the pure gauge theory when quarks are coupled in, because adjoint quarks carry
zero center charge. We confirm these features in our results.

The paper is organized as follows. In Sec.~\ref{Sec:setup} we introduce our
model, working out separately the actions in the gauge and quark sectors. The
gauge part is well known from literature, and we therefore just elaborate on
the Weiss mean-field approximation used in this paper. In the quark part we
deal with the task to construct an interaction Lagrangian with $\gr{SU}(2N_f)$
flavor symmetry. While this was previously achieved for QC$_2$D and actually
applies equally well to all type-II theories, here we construct analogously a
model Lagrangian for type-I theories. Section \ref{Sec:2c} is devoted to
two-color QCD. We study the phase diagram of aQC$_2$D and derive the
Ginzburg--Landau (GL) theory that governs the behavior of the system near the
tetracritical point. We find a simple \emph{closed} analytic expression for the
expectation values of the Polyakov loop in all representations, valid in pure
gauge theory as well as with dynamical quarks in an arbitrary representation.
In Sec.~\ref{Sec:3c} we show analogous results for aQCD. Finally, in
Sec.~\ref{Sec:conclusions} we summarize and conclude. Some technical details
are deferred to the appendices. Throughout the paper, we use the natural units
in which the Planck's and Boltzmann's constants as well as the speed of light
are equal to one, and the timelike metric in the Minkowski space.


\section{Model setup}
\label{Sec:setup} In this section we set up the model that we later on use for
numerical computations. In the gauge sector we employ a simple lattice-inspired
model, which can in principle be used for any number of colors. The quark NJL
Lagrangian derived afterwards is, as already stressed, applicable to all
QCD-like theories with quarks in a real representation. This is natural: the
Lagrangian is based almost exclusively on the flavor symmetry and is therefore
valid for an arbitrary number of colors. The numerical values of the parameters
in the model will be fixed in the following sections when we come to the
discussion of concrete results.


\subsection{Gauge sector}
\label{Subsec:setupgauge} Our starting point for the pure gauge sector is an
effective theory for the Polyakov loop inspired by the lattice strong-coupling
expansion. We closely follow the notation and line of argument of
Ref.~\cite{Abuki:2009gauge}. The action of the model is given by
\begin{equation}
\Sa_g[L]=-N_c^2e^{-a/T}
\sum_{\bs x,\bs y}\ell_{\rm F}(\bs{x})\ell^*_{\rm F}(\bs x+\bs y)\,,
\label{eq:sgauge}
\end{equation}
where $\bs x$ are the lattice sites and $\bs y$ are the neighboring sites. (We
use boldface to indicate spatial vectors.) The only adjustable parameter $a$ is
related to the string tension and can be extracted from numerical simulations
of the full (pure) gauge theory. Furthermore, $\ell_{\rm F}(\bs
x)\equiv\frac{1}{N_c}\tr L_{\rm F}(\bs x) $ is the traced Polyakov loop in the
fundamental representation; in the full gauge theory, the Polyakov loop in a
given representation $\mathcal{R}$ is defined as
\begin{equation}
L_\mathcal{R}(\bs x)\equiv\mathcal{P}\exp\biggl[\imag\int_0^{1/T}
\dd\tau\, A_4^a(\bs x,\tau)T_{a\mathcal{R}}\biggr]\,,
\label{polyakovgeneral}
\end{equation}
where $T_{a\mathcal{R}}$ are the gauge generators in this representation.

In the so-called Polyakov gauge where temporal gluons have constant
values, this simplifies to
\begin{equation}
L_\mathcal{R}(\bs x)=\exp\left[\imag A_4^a(\bs x)T_{a\mathcal{R}}/T\right].
\end{equation}
Moreover, only the components of $A_4^a$ corresponding to generators that form
the Cartan subalgebra of the gauge group are nonzero. Let these components be
$\theta_iT$. (There are $N_c-1$ independent ones; the conventional factor $T$
makes the variables $\theta_i$ dimensionless.)  Each representation of the gauge
group is characterized by a set of weights, $w_{i\alpha}$, that represent the
eigenvalues of the generators of the Cartan subalgebra in this representation;
the index $\alpha$ labels the different eigenvectors of the Cartan subalgebra.
The traced Polyakov loop in representation $\mathcal R$ then reads
\begin{equation}
\ell_{\mathcal R}(\bs x)=\frac1{\dim\mathcal R}\sum_\alpha e^{\imag\theta_i
(\bs x)w_{i\alpha}}\,.
\end{equation}
In the fundamental representation, the Polyakov loop (in the Polyakov gauge) is
usually represented as
$\mathrm{diag}(e^{\imag\theta_1},\dotsc,e^{\imag\theta_{N_c-1}},
e^{-\imag(\theta_1+\dotsb+\theta_{N_c-1})})$. This corresponds to the choice of
the $N_c$ weights of the fundamental representation as
$w_{i\alpha}=\delta_{i\alpha}$ for $\alpha=1,\dotsc,N_c-1$, and $w_{iN_c}=-1$
for all $i$. Equivalently, it can be written by defining
$\theta_{N_c}=-(\theta_1+\dotsb+\theta_{N_c-1})$ up to an integer multiple of
$2\pi$.

In the Weiss mean-field approximation, the nearest-neighbor interaction is
linearized and the action \eqref{eq:sgauge} is replaced with the action
$\Sa_{\text{mf}}(\alpha,\beta)$, depending on two mean fields
$\alpha,\beta$,\footnote{Here, we adhere to the notation introduced in
Ref.~\cite{Abuki:2009gauge}. Let us therefore just stress that the symbol
$\beta$ is not to be confused with the inverse temperature.}
\begin{equation}
\Sa_{\text{mf}}(\alpha,\beta)=
-N_c\sum_{\bs x}[\alpha\re\ell_{\rm F}(\bs x)+\imag\beta\im\ell_{\rm F}(\bs x)]\,.
\label{generalMFaction}
\end{equation}
The dynamical variables of the model \eqref{eq:sgauge} are the (untraced)
Polyakov loops $L(\bs x)$ and its partition function is therefore obtained as
$\Za_g\equiv\exp(-\Omega_g/T)=\int\prod_{\bs x}\dd L(\bs x)\exp(-\Sa_g[L])$,
where $\dd L$ is the group-invariant (Haar) measure on the $\gr{SU}(N_c)$ gauge
group. For the sake of future reference, let us add that in terms of the phases
$\theta_i$, the Haar measure can be written as
\begin{equation}
\dd
L=\prod_{i=1}^{N_c-1}\dd\theta_i\prod_{i<j}^{N_c}
|e^{\imag\theta_i}-e^{\imag\theta_j}|^2\,,
\label{haar}
\end{equation}
The integration over the variables $\theta_i$ is performed over the range
$[0,2\pi]$.

The thermodynamic potential can now be rewritten by subtracting and adding the
mean-field action, resulting in the expression
\begin{equation}
\frac{\Omega_g}{T}=-\log\bigl\langle
e^{-(\Sa_g-\Sa_{\text{mf}})}\bigr\rangle_{\text{mf}}-
\log\int\prod_{\bs x}\dd L(\bs x)\,e^{-\Sa_{\text{mf}}}\,.
\label{intermediate}
\end{equation}
Here and in the following, $\mfave\cdot$ is the average
with respect to the distribution defined by the mean-field action. For a given
(not necessarily local) function $\mathcal O[L]$ of the Polyakov loop, it reads
\begin{equation}
\mfave{\mathcal O}=\frac{\displaystyle\int\prod_{\bs x}\dd L(\bs x)\,
\mathcal O[L]e^{-\Sa_{\text{mf}}}}
{\displaystyle\int\prod_{\bs x}\dd L(\bs x)\,e^{-\Sa_{\text{mf}}}}\,.
\label{groupaverage}
\end{equation}
Note that when the function $\mathcal O$ is local and does not depend
explicitly on the coordinate, the product over lattice sites can be dropped.

Equation~\eqref{intermediate} is still exact; no approximation has been made so
far. By the same token, the thermodynamic potential $\Omega_g$ is independent
of the arbitrary variables $\alpha,\beta$. In the Weiss mean-field
approximation, one replaces $\bigl\langle
e^{-(\Sa_g-\Sa_{\text{mf}})}\bigr\rangle_{\text{mf}}$ with
$e^{-\mfave{\Sa_g-\Sa_{\text{mf}}}}$ \cite{Abuki:2009gauge}. The mean fields
are then determined selfconsistently from the stationarity condition. In fact,
as long as $\beta=0$ so that the averaging is done with a real mean-field
action, one can use Jensen's inequality\footnote{Jensen's inequality (see
e.g.~Ref.~\cite{Lohwater:1982lo}) states rather generally that for any real
convex function $f$, $f(\langle x\rangle)\leq \langle f(x)\rangle$, where the
averaging involves either a (weighted) arithmetic mean in the discrete version
of the inequality, or an integral average over a given probability distribution
in the continuous version.} to show that this approximation provides a strict
upper bound for the exact free energy. Its optimum estimate is then obtained by
minimizing with respect to $\alpha$.

The final formula for the Weiss mean field gauge thermodynamic potential reads
\begin{equation}
\begin{split}
\frac{\Omega^{\rm W}_ga_s^3}{TV}=&-2(d-1)N_c^2e^{-a/T}\mfave{\ell_{\rm F}}
\mfave{\ell^*_{\rm F}}+\frac{N_c}{2}\bigl[(\alpha+\beta)\mfave{\ell_{\rm F}}+(\alpha-\beta)
\mfave{\ell^*_{\rm F}}\bigr]-\\
&-\log\int\dd L\,e^{N_c(\alpha\re\ell_{\rm F}+\imag\beta\im\ell_{\rm F})}\,.
\end{split}
\label{weiss}
\end{equation}
Here $a_s$ denotes the lattice spacing and the factor $a_s^3/V$ is just the
inverse of the number of lattice sites; $d$ stands for the dimensionality of
spacetime so that $2(d-1)$ is the number of nearest neighbors on a cubic
lattice.


\subsection{Quark sector}
\label{Subsec:setupNJL}

The Lagrangian of the quark sector cannot be derived from the underlying gauge
theory directly. However, it is strongly constrained by the requirement that it
inherits all the symmetries of the QCD-like theory. As already stressed above,
in theories with $N_f$ massless quark flavors in a (pseudo)real representation
of the gauge group, the usual chiral symmetry is promoted to $\gr{SU}(2N_f)$.
In order to see how this comes about, let us start from the Lagrangian of the
gauge theory, including a common mass $m_0$ for all quark flavors,
\begin{equation}
\La_{\text{QCD-like}}=\psib\imag\slashed\Da\psi-m_0\psib\psi\,,
\label{QCDlikeLagr}
\end{equation}
where $\Da_\mu\psi=(\partial_\mu-\imag gT_aA^a_\mu)\psi$ is the gauge-covariant
derivative. Indices are suppressed so that this formula holds for quarks in any
representation of the gauge group.

The fact that the quark representation is (pseudo)real means that there is a
unitary matrix $\Pa$ such that $\Pa\psic$ has the same transformation
properties under the gauge group as $\psi$. It is then advantageous to trade
the Dirac spinor, consisting of the left- and right-handed components, for the
purely left-handed Nambu spinor,
\begin{equation}
\Psi=\begin{pmatrix}
\psi_{\rm L}\\
\Pa\psic_{\rm R}
\end{pmatrix}.
\label{Nambu}
\end{equation}
A crucial fact known from the theory of Lie algebras is that $\Pa$ is either
symmetric or antisymmetric according to whether the quark representation is
real or pseudoreal~\cite{Georgi:1982jb}. Writing collectively $\Pa^T=\pm\Pa$,
we can introduce the charge-conjugated Nambu spinor,
\begin{equation}
\Psic=\Pa\begin{pmatrix}
\psic_{\rm L}\\
(\Pa\psic_{\rm R})^\cc
\end{pmatrix}=
\begin{pmatrix}
\Pa\psic_{\rm L}\\
\pm\psi_{\rm R}
\end{pmatrix}.
\end{equation}
The Dirac conjugate of both $\Psi$ and $\Psic$ is defined naturally by
conjugating the individual components. The Lagrangian \eqref{QCDlikeLagr}
then becomes, in the Nambu formalism,
\begin{equation}
\La_{\text{QCD-like}}=\Psib\imag\slashed\Da\Psi-\left[
\frac12m_0\Psicb
\begin{pmatrix}
0 & \openone\\
\pm\openone & 0
\end{pmatrix}
\Psi+\hc
\right].
\label{QCDlikeLagrNambu}
\end{equation}
First of all, we can see that in the chiral limit, the Lagrangian of a QCD-like
theory indeed has an $\gr{SU}(2N_f)$ symmetry. Note that baryon number is
already incorporated in this simple group, for it is represented by the block
matrix $\frac12\mathrm{diag}(\openone,-\openone)$ in Nambu space. The change of
the overall phase of the Nambu spinor corresponds to the axial $\gr{U(1)_A}$
symmetry which is broken at the quantum level by instanton effects. Since the
mass term has the same structure as the chiral condensate, we can also
immediately infer that for type-I (type-II) theories the order parameter for
flavor symmetry breaking transforms as a(n) (anti)symmetric rank-two tensor of
$\gr{SU}(2N_f)$. Therefore, the two classes of theories have different
symmetry-breaking patterns and subsequently also different low-energy spectra.
The symmetry-breaking patterns in the vacuum are
$\gr{SU}(2N_f)\to\gr{SO}(2N_f)$ and $\gr{SU}(2N_f)\to\gr{Sp}(2N_f)$ for type~I
and type~II, respectively~\cite{Kogut:2000qcdlike}.

The task to construct an NJL-type interaction compatible with the
$\gr{SU}(2N_f)$ symmetry is most easily accomplished using the Nambu notation
\eqref{Nambu}. It is useful to stress right at the outset that as long as only
color-singlet channels are considered, each of the Lagrangians to be
constructed below applies to the whole class of QCD-like theories (type-I or
type-II), regardless of the detailed structure of the gauge group or the quark
representation. In fact, NJL Lagrangians for type-II theories with two quark
flavors were already constructed in Ref.~\cite{Andersen:2010phase}. Here we
follow the same line of argument with the necessary modifications for the type-I
case.

One property that further distinguishes the type-I and type-II theories is the
severity of the sign problem. While we remarked before that all QCD-like
theories considered in this paper are free from the sign problem, one should be
a bit careful with the type-II theories. There, the determinant of the Dirac
operator is in general real, but needs not be positive. In order that there be
no sign problem, one therefore has to consider an even number of flavors. On
the other hand, type-I theories have no sign problem for any number of
flavors~\cite{Hands:2000numerical}. As a warm-up exercise, we thus start with
the simplest case of one flavor.

In the following, the Pauli matrices
$\sigma_{0,1,2,3}=\{\openone,\sigma_1,\sigma_2,\sigma_3\}$ are used
to denote the block matrices in Nambu space, and
$\tau_{0,1,2,3}=\{\openone,\tau_1,\tau_2,\tau_3\} $ are used to denote the
flavor generators for $N_f=2$. The symmetric rank-two tensor representation of
the flavor $\gr{SU(2)\simeq SO(3)}$ group is real and three-dimensional. Using
the basis of symmetric unimodular unitary matrices as
$\vec\Sigma=\{\openone,\imag\sigma_1,\imag\sigma_3\}$, we can immediately
construct two
four-fermion interaction terms,
\begin{equation}
\begin{split}
\La_{\gr{1f,U(2)}}&=G\bigl|\Psicb\vec\Sigma\Psi\bigr|^2
=G\bigl[(\psib\psi)^2+(\psib
\imag\gamma_5\psi)^2+|\psicb\gamma_5\psi|^2+|\psicb\psi|^2\bigr]\,,\\
\La_{\gr{1f,SU(2)}}&=
\frac{G}{2}\Bigl[\bigl(\Psicb\vec\Sigma\Psi\bigr)^2+\hc\Bigr]
=-G\bigl[(\psib\psi)^2-(\psib
\imag\gamma_5\psi)^2+|\psicb\gamma_5\psi|^2-|\psicb\psi|^2\bigr]\,.
\end{split}
\label{eq:la1fsu2}
\end{equation}
While the former preserves the axial $\gr{U(1)_A}$, the latter breaks it
explicitly. It is easy to verify that $\La_{\gr{1f,SU(2)}}$ is the 't~Hooft
determinant term, i.e.
\begin{equation}
\La_{\gr{1f,SU(2)}}=2G(\det \overline{\Psic_i}\Psi_j +\hc)\,.
\end{equation}

For two flavors, the ten basis matrices of the symmetric rank-two tensor
representation of the flavor $\gr{SU}(4)$ are chosen as the symmetric Kronecker
products of $\sigma$ and $\tau$, i.e.
\begin{equation}
\vec\Sigma=\{
\sigma_{\text{sym}}\otimes\tau_{\text{sym}},\sigma_{\text{antisym}}\otimes
\tau_{\text{antisym}}\}\,.
\end{equation}
Since the 10-dimensional representation of $\gr{SU(4)}$ is complex, only one of
the above two possibilities to construct an invariant interaction term remains,
\begin{equation}
\begin{split}
\La_{\gr{2f,U(4)}}&=G\bigl|\Psicb\vec\Sigma\Psi\bigr|^2\\
&=G\Bigl[(\bar{\psi}\psi)^2+(\bar{\psi}\imag\gamma_5\vec\tau\psi)^2
+(\bar{\psi}\imag\gamma_5\psi)^2+(\bar{\psi}\vec{\tau}\psi)^2
+\sum_S|\psicb\tau_S\psi |^2+\sum_S|\psicb\gamma_5
\tau_S\psi|^2\Bigr]\,,
\end{split}
\label{eq:la2fu2}
\end{equation}
which preserves $\gr{U(1)_A}$ automatically. (Here $\tau_S$ denotes the set of
symmetric Pauli matrices, $\tau_S=\{\openone,\tau_1,\tau_3\}$.) A $\gr{U(1)_A}$
breaking interaction can again be introduced by the 't~Hooft determinant term,
but such a term will be an eight-fermion contact interaction which we do not
consider in our model.


\subsection{Mean-field approximation}
\label{Subsec:setupMF} We will employ the usual mean-field approximation,
introducing the collective bosonic fields via the Hubbard--Stratonovich
transformation and subsequently replacing them with their vacuum expectation
values. To that end, however, one first needs to guess which condensates (order
parameters) will appear in the phase diagram. The case of type-II theories with
two quark flavors was worked out in Ref.~\cite{Andersen:2010phase}: as long as
just the baryon chemical potential is considered, one only needs the chiral
condensate, $\sigma=-2G\langle\psib\psi\rangle$, and the scalar diquark
condensate, $\Delta=2\imag G\langle\psi^TC\gamma_5\Pa\tau_2\psi\rangle$. Since
the diquark wave function is antisymmetric in color as well as spin indices, it
must, by means of the Pauli principle, also be antisymmetric with respect to
flavor. The (spin-zero) diquark in type-II theories therefore mixes quarks of
different flavors. Consequently, in the presence of an isospin chemical
potential the diquark pairing feels stress and eventually diminishes via a
first-order phase transition, with a narrow window of chemical potentials
featuring inhomogeneous pairing~\cite{Andersen:2010phase,Fukushima:2007bj}.

In type-I theories the scalar order parameters are symmetric in color and
antisymmetric in spin indices, hence they must be symmetric in flavor. This is
in accordance with the fact that for two flavors, there are altogether nine NG
bosons of the $\gr{SU(4)/SO(4)}$ coset, the isospin triplet of pions and the
isospin triplet of (complex) diquarks. At zero isospin chemical potential, the
isospin multiplets are strictly degenerate. In particular all $uu$, $dd$, and
$ud+du$ diquarks can condense when the baryon chemical potential exceeds their
common mass. However, for arbitrarily small isospin chemical potential, the
diquarks formed from quarks of the same flavor will be favored. Such
single-flavor condensates do not feel stress at nonzero chemical potential, and
the phase diagram of type-I theories will therefore not contain inhomogeneous
phases, as observed in Ref.~\cite{Splittorff:2000mm}.

With the above argument in mind, we restrict our attention to single-flavor
condensates. The fact that the two-flavor four-fermion interaction
\eqref{eq:la2fu2} automatically preserves $\gr{U(1)_A}$ means that the
condensates differing just by opposite parity will be degenerate. However, we
know from the Vafa--Witten theorem that in the vacuum parity is not
spontaneously broken~\cite{Vafa:1984xg}. The degeneracy will be eventually
lifted by instanton effects, manifested in the eight-quark `t~Hooft interaction
term. Within the present model, we will simply ignore the negative-parity channels.

As long as we only deal with one-flavor condensates, we can write down the
contribution to the thermodynamic potential from a single quark flavor. This
equals the thermodynamic potential of free fermionic quasiparticles. In
presence of a pairing gap $\Delta$, their dispersion relation reads $E^e_{\bs
k}=\sqrt{(\xi^e_{\bs k})^2+\Delta^2}$, where $\xi^e_{\bs k}=\epsilon_{\bs
k}+e\mu$, $e=\pm$, and $\epsilon_{\bs k}=\sqrt{\bs k^2+M^2}$; $M=m_0+\sigma$ is
the constituent quark mass and $\mu$ the quark chemical potential. The gauge and
quark sectors are coupled in the PNJL spirit \cite{Fukushima:2004chiral}. In the
Polyakov gauge the temporal component of the gauge field is constant. The
individual quark color states in a given representation will then have, in the
presence of the background gauge field, effective chemical potentials $\imag
T\sum_i\theta_iw_{i\alpha}$. Since the quasiparticle spectrum discussed above is
the same for all color states in the representation (this is because all
condensates are color singlets!), the thermodynamic potential of one quark
flavor will simply be
\begin{equation}
\begin{split}
\frac{\Omega_q}{VN_f}=&\frac{\sigma^2+\Delta^2}{4G}-\sum_e
\threeint{\bs k}\times\\
&\times\Bigl\{E^e_{\bs k}\dim\mathcal R+2T
\log\Bigl\langle\prod_\alpha\left[1+2\cos(\theta_iw_{i\alpha}) e^{-E^e_{\bs
k}/T}+e^{-2E^e_{\bs k}/T}\right]^{1/2}\Bigr\rangle_{\text{mf}}\Bigr\}\,.
\end{split}
\label{Omega1flavor}
\end{equation}
The power of $1/2$ in the second line compensates the doubling of the number of
degrees of freedom in the Nambu formalism.

The group average must be performed once we couple the quarks to the Polyakov
loop. Note that we do not average the full quark thermodynamic potential, but
only the argument of the logarithm. This replacement was introduced in Eq.~(13)
of Ref.~\cite{Abuki:2009gauge} as a convenient approximation to
$\mfave{\Omega_q}$. However, in Appendix \ref{App:averaging} we present a
heuristic argument showing that the prescription \eqref{Omega1flavor} is
actually superior to the full average $\mfave{\Omega_q}$. While with
fundamental quarks considered in Ref.~\cite{Abuki:2009gauge} the numerical
difference between the two ways of evaluating the quark sector thermodynamic
potential is negligible, we point out that with adjoint quarks, taking the
average $\mfave{\Omega_q}$ would lead to unphysical artifacts which are not
present in Eq.~\eqref{Omega1flavor}.


\subsection{Parameter fixing in the quark sector}
\label{Subsec:setupparfix} The NJL part of the model has three adjustable
parameters: the coupling $G$, the current quark mass $m_0$, and the ultraviolet
cutoff that regulates divergent integrals. (Within this paper, we will use the
three-momentum regularization scheme.) These need to be fixed by fitting to
three selected observables. A conventional, and convenient, choice are the
chiral condensate, pion mass, and decay constant in the vacuum. While the pion
mass is more or less a free parameter that can be easily modified in lattice
simulations by tuning the quark mass, the remaining two parameters depend on
the single physical scale of the underlying theory, and cannot therefore be
adjusted at will.

In three-color QCD with fundamental quarks, one can directly use experimental
observables. In QC$_2$D, the input parameters were determined in
Ref.~\cite{Brauner:2009twocolor} from their three-color counterparts by
$N_c$-rescaling. Unfortunately, we are not aware of suitable lattice data that
would allow us to fix the parameters directly in the case of aQCD and aQC$_2$D.
We therefore use the following indirect argument. Suppose that we have a theory
with both fundamental and adjoint quarks. Gauge invariance can then only be
maintained when the coupling of quarks to gluons is the same in both
representations. Since the effective meson-channel Lagrangians of the NJL type
can be derived from a one-gluon-exchange-inspired interaction, this allows us
to fix the ratio of the effective couplings in the fundamental and adjoint
quarks sectors.

Concretely, assume the current--current interaction
\begin{equation}
\La_{\text{int}}=-g(\psib\gamma^\mu T_{a\mathcal R}\psi)^2\,.
\label{current}
\end{equation}
The coupling $g$ can be directly related to the microscopic QCD coupling and
the screening mass of the gluon in the one-gluon-exchange approximation. We
therefore assume that it is the same for fundamental and adjoint quarks.
Performing the Fierz transformation to the meson channel yields
the effective NJL coupling $G_{\text F}=g(N_c^2-1)/(2N_c^2N_f)$ for fundamental
quarks \cite{Buballa:2003qv}. For adjoint quarks we analogously obtain
$G_{\text A}=gN_c/[(N_c^2-1)N_f]$. This results in the ratio
\begin{equation}
\frac{G_{\text{A}}}{G_{\text{F}}}=\frac{2N_c^3}{(N_c^2-1)^2}\,.
\label{GAGFratio}
\end{equation}
For the reader's convenience, the derivation of this relation is sketched in
Appendix \ref{App:Fierz}. In the following sections, we will use it to infer the
value of the coupling for adjoint quarks from that for the fundamental ones. We
will not refer to the original current--current interaction anymore.

Eq.~\eqref{GAGFratio} would at first glance suggest that the coupling for
adjoint quarks is weaker than for the fundamental ones (with the exception
$N_c=2$). One may then wonder why the chiral restoration temperature is much
higher for adjoint quarks. The reason for this is that in the gap equation, the
coupling is multiplied by the number of quark degrees of freedom coming from
the quark loop. The effective coupling ratio for adjoint versus fundamental
quarks therefore is $2N_c^2/(N_c^2-1)$ which is always larger than two.


\section{Two colors}
\label{Sec:2c} For two colors, the group integration is easily done and it is
possible to find closed analytic expressions for all general formulas derived
above. First, there is just one independent phase $\theta$, associated with the
only diagonal generator of the $\gr{SU(2)}$ gauge group. The
$(2j+1)$-dimensional spin-$j$ representation then has weights
$-2j\theta,\dotsb,+2j\theta$, and one immediately obtains
\begin{equation}
\ell_j=\frac{1}{2j+1}\frac{\sin(2j+1)\theta}{\sin\theta}\,.
\label{2colorpolyakov}
\end{equation}
The Haar measure \eqref{haar} reduces to $\dd
L=\frac1\pi\sin^2\theta\,\dd\theta$, normalized so that the group volume is
unity.
\TABLE[t]{%
\setlength{\tabcolsep}{0.7em}
\renewcommand{\arraystretch}{1.5}
\begin{tabular}{ccccc}
\hline\hline $a\text{ [MeV]}$ & $b^{1/3}\text{ [MeV]}$ & $\Lambda\text{ [MeV]}$
&
$G\text{ [GeV$^{-2}$]}$ & $m_0\text{ [MeV]}$\\
\hline
$670.9$ & $269.2$ & $657$ & $25.71$ & $5.4$\\
\hline\hline
\end{tabular}
\caption{Model parameters for two-color QCD with adjoint quarks.}
\label{Tab:2colorpars}%
}

Since all traced Polyakov loops of $\gr{SU(2)}$ are real, we need just one mean
field $\alpha$ in Eq.~\eqref{generalMFaction}. Using the definition of the
modified Bessel function of integer order,
\begin{equation}
I_n(x)=\frac1\pi\int_0^\pi\dd\theta\,e^{x\cos\theta}\cos n\theta\,,
\label{bessel}
\end{equation}
and the recurrence relation $I_{n-1}(x)-I_{n+1}(x)=\frac{2n}xI_n(x)$, one
derives the expectation value of the Polyakov loops \cite{Ambjorn:1984mb},
\begin{equation}
\mfave{\ell_j}=\frac{I_{2j+1}(2\alpha)}{I_1(2\alpha)}\,.
\label{2colorpolyakovave}
\end{equation}
The gauge part of the thermodynamic potential \eqref{weiss} in turn becomes
\begin{equation}
\frac{\Omega^{\rm W}_g}V=bT\left[
-24e^{-a/T}\mfave{\ell_{\rm F}}^2+2\alpha\mfave{\ell_{\rm F}}-
\log\frac{I_1(2\alpha)}\alpha\right],
\label{2colorOmegag}
\end{equation}
where we denoted $b=a_s^{-3}$ to facilitate comparison with the ``standard''
PNJL model \cite{Brauner:2009twocolor,Abuki:2009gauge}. The weights of the
adjoint representation are $-2,0,2$ and the group average in the quark sector is
also easily evaluated. The result is most conveniently written in terms of the
expectation value of the adjoint Polyakov loop,
\begin{equation}
\begin{split}
\frac{\Omega_q}{VN_f}=&\frac{\sigma^2+\Delta^2}{4G}-\sum_e\threeint{\bs k}
\Bigl[3E^e_{\bs k}+2T\log\bigl(1+e^{-E^e_{\bs k}/T}\bigr)+\\
&+2T\log\bigl(1-e^{-E^e_{\bs k}/T}+e^{-2E^e_{\bs k}/T}
+3e^{-E^e_{\bs k}/T}\mfave{\ell_{\rm A}}\bigr)\Bigr]\,.
\end{split}
\label{2colorOmegaq}
\end{equation}
This is the formula that we use for the analysis of the phase diagram.


\subsection{Phase diagram}
\label{Subsec:2cPD}
\FIGURE[l]{%
\includegraphics[scale=0.85]{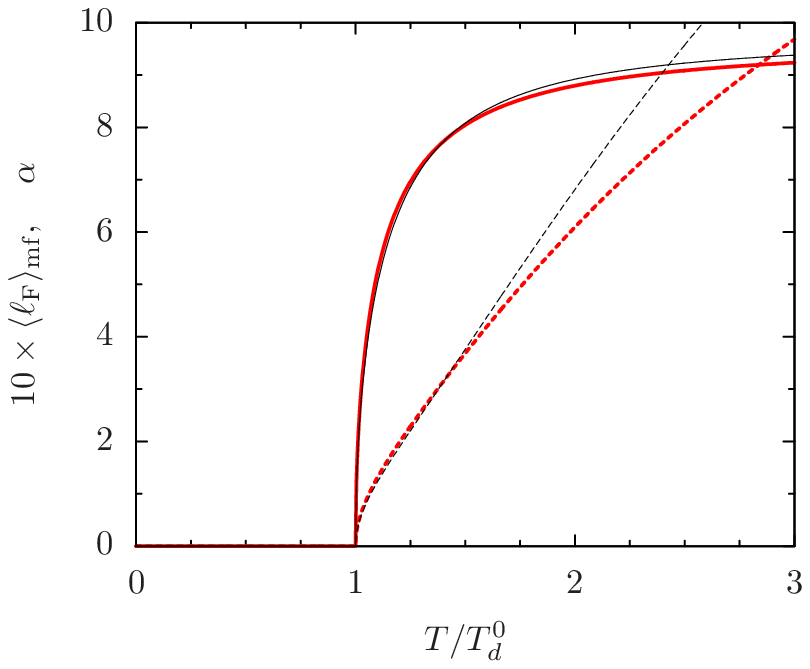}
\caption{Comparison of the expectation values of the mean field $\alpha$
(dashed) and the fundamental Polyakov loop (solid) in the naive (thin black
lines) and Weiss (thick red lines) mean-field approximations to the pure gauge
theory.}
\label{Fig:pure2color}%
}
The first thing that needs to be done is to fix the parameters of our model.
There are altogether five of them: the coupling, current quark mass, and cutoff
in the quark sector, and $a,b$ in the gauge sector. Our method to estimate the
NJL input parameters was explained in Sec.~\ref{Subsec:setupparfix}, so we
simply use the parameter set for QC$_2$D established in
Ref.~\cite{Brauner:2009twocolor} and rescale the coupling
according to Eq.~\eqref{GAGFratio}. Also, we introduce an additional factor of
two to account for the fact that we deal with one quark flavor only here. As to
the gauge sector, we use the same \emph{physical} input as in
Ref.~\cite{Brauner:2009twocolor}, that is, critical temperature in the pure
gauge theory $T_d^0=270\text{ MeV}$ and the string tension $\sigma_s=(425\text{
MeV})^2$. These values were obtained from the three-color pure gauge theory
using their scaling properties in the limit of a large number of colors, so
they do not quantitatively precisely agree with those one would obtain directly
from the two-color lattice gauge theory. However, this does not matter since we
do not fit the parameters in the quark sector to lattice data. Within this
paper, we merely wish to demonstrate the general trends as the number of
colors or the quark representation are varied.

Since we use a different potential for the Polyakov loop than in
Ref.~\cite{Brauner:2009twocolor}, the parameters $a,b$ will actually take
different values despite the same input for $T_d^0$ and $\sigma_s$. The
deconfinement transition in the pure gauge theory is of second order with two
colors, hence we can expand the thermodynamic potential \eqref{2colorOmegag} to
second order in $\alpha$,
\begin{equation}
\frac{\Omega^{\rm W}_g}{V}=bT\alpha^2\left(\frac12-6e^{-a/T}\right)+\mathcal
O(\alpha^4)\,.
\end{equation}
From here one concludes that $a=T_d^0\log12$. The lattice spacing $a_s$, hence
the parameter $b$, is then determined from the strong-coupling relation
$a=\sigma_sa_s$. The numerical values of all parameters are summarized in
Tab.~\ref{Tab:2colorpars}.

The Weiss mean-field approximation employed here differs from the mean-field
approximation used in Ref.~\cite{Brauner:2009twocolor}, which we will
henceforth refer to as ``naive'' for reasons explained in Appendix
\ref{App:averaging}. In the latter, the gauge sector potential can be expressed
solely in terms of the fundamental Polyakov loop and it reads,
\begin{equation}
\frac{\Omega^{\rm naive}_g}V=-bT\bigl[24e^{-a/T}\ell_{\rm F}^2+
\log(1-\ell_{\rm F}^2)\bigr]\,,
\label{2colorOmegagnaive}
\end{equation}
cf.~Eq.~\eqref{2colorOmegag}. It is therefore mandatory to compare the results
obtained with the two approaches. We do so within the pure gauge theory. The
expectation values of the fundamental Polyakov loop and the mean field $\alpha$
are shown in Fig.~\ref{Fig:pure2color}.\footnote{Note that there is no
$\alpha$ in the naive mean-field approximation. The values plotted in
Fig.~\ref{Fig:pure2color} were obtained by inverting the relation
\eqref{2colorpolyakovave}.} It is obvious that the results for the Polyakov
loop are not sensitive to the particular implementation of the gauge sector as
long as the parameters are adjusted to reproduce the same physical observables.

\FIGURE[t]{%
\includegraphics[scale=0.9]{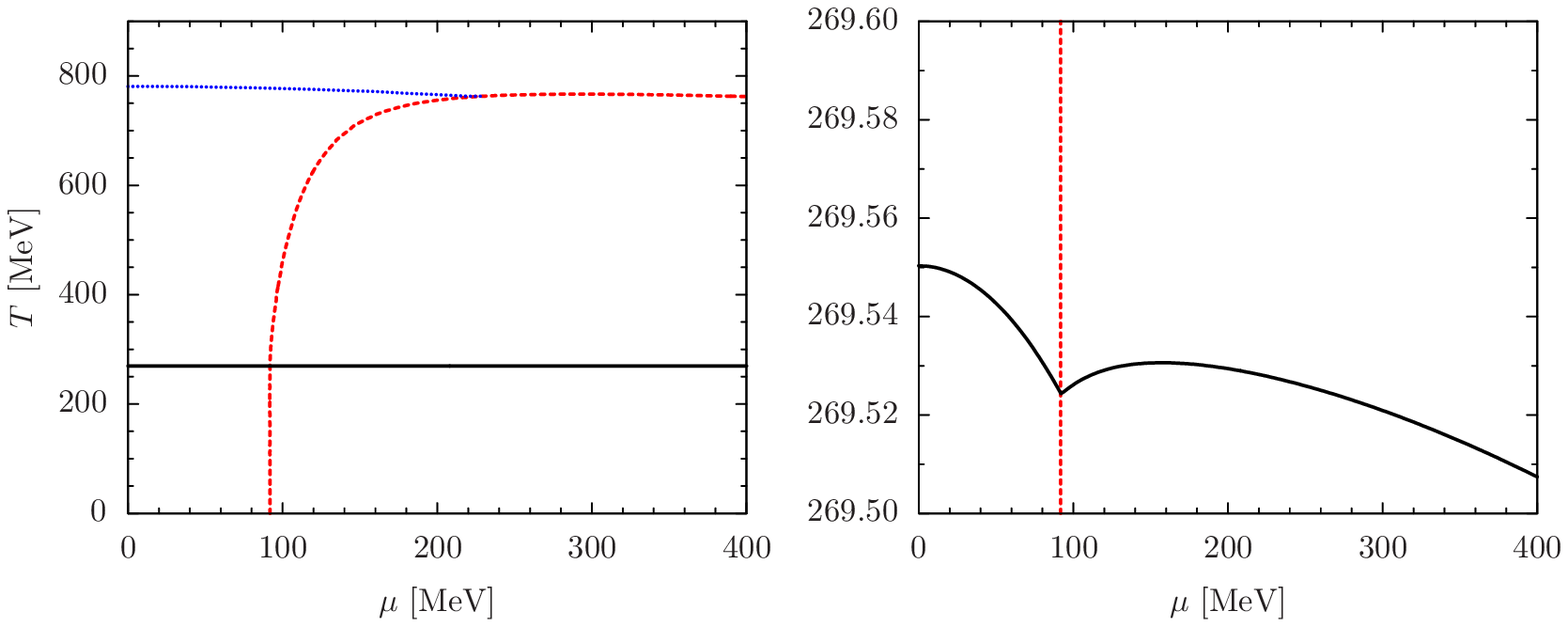}
\caption{Phase diagram of two-color QCD with one flavor of adjoint quarks.
Black solid line: deconfinement transition. Red dashed line: BEC transition.
Blue dotted line: chiral crossover. The right panel zooms in the temperature
scale so that the cusp in the deconfinement critical line is visible.}
\label{Fig:PD2color}%
}
Figure \ref{Fig:PD2color} shows the phase diagram of aQC$_2$D with one quark
flavor in the plane of temperature and quark chemical potential. The
deconfinement transition associated with the breaking of center $\gr Z_2$ is
denoted by the black solid line, while the BEC transition at which the baryon
number $\gr{U(1)_B}$ is broken is indicated by the red dashed line. In addition
to these two sharp phase transitions, there is a smooth crossover associated
with the melting of the chiral condensate. Its position, shown in the left
panel of Fig.~\ref{Fig:PD2color} by the blue dotted line, is defined here by the
maximum temperature gradient of $\sigma$. In the chiral limit, this also becomes
a sharp second-order phase transition. As expected, it does appear at a
temperature much higher than that of the deconfining transition ($T_d=270\text{
MeV}$, while $T_\chi=780\text{ MeV}$ so that $T_\chi/T_d=2.89$). However, the
precise value of this temperature as determined by our model is strongly
affected by the cutoff, as is discussed in more detail in
Sec.~\ref{Subsec:3c_PD}.

\FIGURE[r]{%
\includegraphics[scale=0.85]{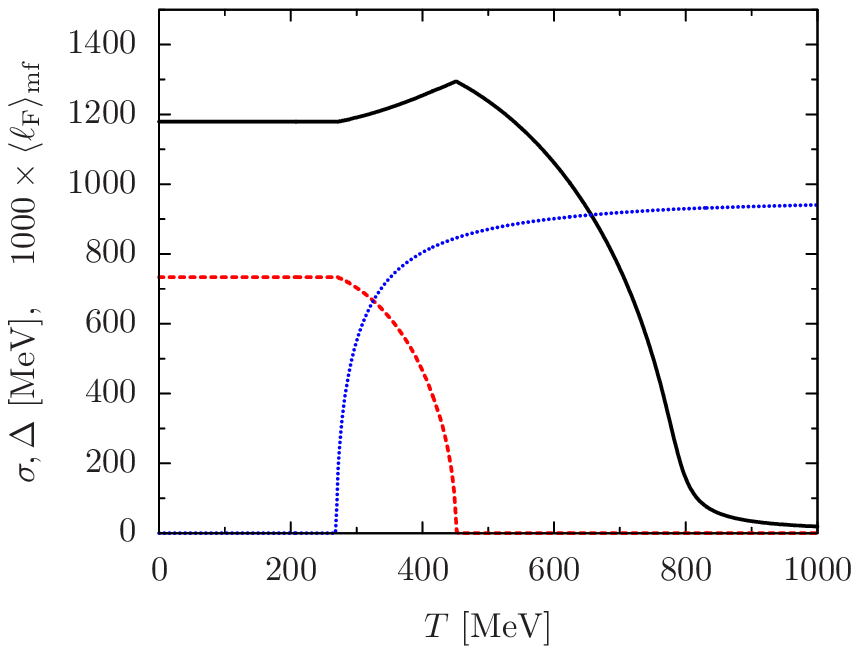}
\caption{Condensates in aQC$_2$D at $\mu=100\text{ MeV}$ as a function of
temperature. The chiral condensate $\sigma$ (black solid line), diquark
condensate $\Delta$ (red dashed line), and the fundamental Polyakov loop (blue
dotted line) are shown.}
\label{Fig:2cconds}%
}
The temperature of the deconfining transition depends on the chemical
potential extremely weakly, even less than in QC$_2$D
\cite{Brauner:2009twocolor}. The reason apparently is that the adjoint quarks
are neutral with respect to the center symmetry. The behavior of the transition
lines in the vicinity of their ``intersection'' will be analyzed in detail in
the following subsection. Finally, the BEC transition at zero temperature
occurs at $\mu=92\text{ MeV}$, which is in a good agreement with the fact that
the mass of the pion/diquark multiplet in the vacuum is $m_\pi=184\text{ MeV}$
within our parameter set.

As an illustration of the solution of the gap equations, we plot in
Fig.~\ref{Fig:2cconds} the condensates at $\mu=100\text{ MeV}$ as a function of
temperature. One can clearly see the effect of the suppression of thermal quark
fluctuations in the confined phase: the condensates are nearly constant for
$T<T_d$.


\subsection{Tetracritical point}
\label{Subsec:2c4critpt}
At the point where the two second-order transition lines cross, the system
exhibits tetracritical behavior \cite{Sannino:2004tetracritical}. Here we will
analyze the details of the phase transitions in the vicinity of the
tetracritical point using the GL theory. The thermodynamic potential depends on
three mean fields, $\alpha,\sigma,\Delta$. Only two of them, $\alpha$ and
$\Delta$, comprise order parameters for spontaneous symmetry breaking of an
exact symmetry (unless we consider the chiral limit). In order to construct the
GL free energy, one therefore needs to eliminate $\sigma$ in favor of
$\alpha,\Delta$ by means of its gap equation. Around the tetracritical point,
we can then perform a double Taylor expansion of the total thermodynamic
potential, $\Omega=\Omega^{\rm W}_g+\Omega_q$. Thanks to the $\gr Z_2$ and
$\gr{U(1)_B}$ symmetries, it depends just on the squares of the mean fields,
\begin{equation}
\frac{\Omega(\alpha^2,\Delta^2)}V=b_\alpha\alpha^2+b_\Delta\Delta^2
+\frac12\left[\lambda_{\alpha\alpha}(\alpha^2)^2
+2\lambda_{\alpha\Delta}\alpha^2\Delta^2+\lambda_{\Delta\Delta}(\Delta^2)^2
\right].
\label{GL}
\end{equation}
The effective quartic couplings are determined by the second \emph{total}
derivatives of the thermodynamic potential,
\begin{equation}
\lambda_{\alpha\alpha}=\frac1V\frac{\dd^2\Omega}{\dd(\alpha^2)^2}\,,\qquad
\lambda_{\alpha\Delta}=\frac1V\frac{\dd^2\Omega}{\dd\alpha^2\dd\Delta^2}\,,
\qquad
\lambda_{\Delta\Delta}=\frac1V\frac{\dd^2\Omega}{\dd(\Delta^2)^2}\,,
\end{equation}
evaluated at $\alpha=\Delta=0$. These total derivatives are in turn given in
terms of the partial derivatives of the thermodynamic potential as a function
of all three mean fields,
\begin{equation}
\frac{\dd^2\Omega}{\dd\chi_i\dd\chi_j}=
\frac{\partial^2\Omega}{\partial\chi_i\partial\chi_j}-
\frac{\partial^2\Omega}{\partial\chi_i\partial\sigma}
\left(\frac{\partial^2\Omega}{\partial\sigma^2}\right)^{-1}
\frac{\partial^2\Omega}{\partial\sigma\partial\chi_j}\,,
\end{equation}
where $\chi_i$ stands for $\alpha^2,\Delta^2$. In order to evaluate the GL
quartic couplings, we need to know six second partial derivatives of the
thermodynamic potential,
\begin{equation}
\begin{split}
\frac{\partial_{\alpha^2\alpha^2}\Omega}V=&\frac14bT\left(16e^{-a/T}-1\right)
+N_fT\sum_e\threeint{\bs k}\frac{\cosh(\xi^e_{\bs k}/T)}
{[2\cosh(\xi^e_{\bs k}/T)-1]^2}\,,\\
\frac{\partial_{\alpha^2\Delta^2}\Omega}V=&N_f\sum_e\threeint{\bs k}
\frac{\sinh(\xi^e_{\bs k}/T)}{\xi^e_{\bs k}}\frac1
{[2\cosh(\xi^e_{\bs k}/T)-1]^2}\,,\\
\frac{\partial_{\Delta^2\Delta^2}\Omega}V=&\frac34N_f\sum_e\threeint{\bs k}
\frac1{(\xi^e_{\bs k})^3}\left[\tanh\frac{3\xi^e_{\bs k}}{2T}-
\frac{3\xi^e_{\bs k}}{2T\cosh^2(3\xi^e_{\bs k}/2T)}
\right],\\
\frac{\partial_{\sigma\alpha^2}\Omega}V=&2MN_f\sum_e\threeint{\bs k}
\frac1{\epsilon_{\bs k}}\frac{\sinh(\xi^e_{\bs k}/T)}
{[2\cosh(\xi^e_{\bs k}/T)-1]^2}\,,\\
\frac{\partial_{\sigma\Delta^2}\Omega}V=&\frac32MN_f\sum_e\threeint{\bs k}
\frac1{\epsilon_{\bs k}(\xi^e_{\bs k})^2}\left[\tanh\frac{3\xi^e_{\bs k}}{2T}-
\frac{3\xi^e_{\bs k}}{2T\cosh^2(3\xi^e_{\bs k}/2T)}
\right],\\
\frac{\partial_{\sigma\sigma}\Omega}V=&\frac{N_f}{2G}\frac{m_0}M+3M^2N_f\sum_e
\threeint{\bs k}\frac{1}{\epsilon_{\bs k}^3}\left[\tanh\frac{3\xi^e_{\bs
k}}{2T}- \frac{3\epsilon_{\bs k}}{2T\cosh^2(3\xi^e_{\bs k}/2T)} \right].
\end{split}
\label{GLcoefficients}
\end{equation}

\FIGURE[t]{%
\includegraphics[scale=0.9]{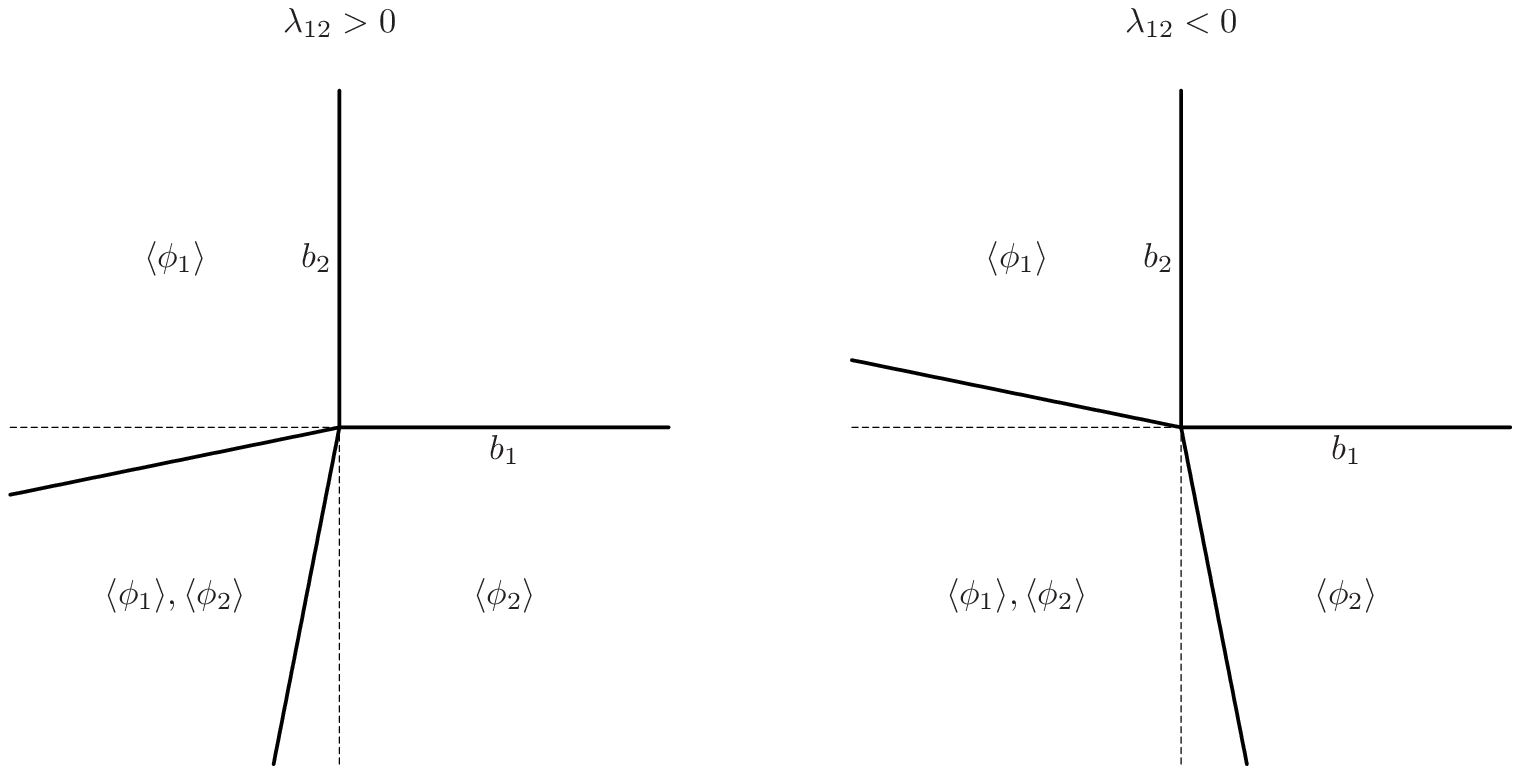}
\caption{Schematic phase diagram of the Ginzburg--Landau theory with two order
parameters. Thick lines denote second-order phase transitions. The labels
indicate which order parameters take nonzero values in a given phase.}
\label{Fig:tetracritical}%
}
In order to see how the two condensates affect each other close to the
tetracritical point, consider the general GL functional with two order
parameters $\phi_{1,2}$ and assume it is constrained to have the form
\begin{equation}
\frac{\Omega(\phi_1,\phi_2)}V=b_1\phi_1^2+b_2\phi_2^2+
\frac12(\lambda_{11}\phi_1^4+2\lambda_{12}\phi_1^2\phi_2^2+\lambda_{22}
\phi_2^4)\,.
\end{equation}
[In our case, all other terms are prohibited by the $\gr Z_2$ and $\gr{U(1)_B}$
symmetries.] The phase diagram of such a model is depicted in
Fig.~\ref{Fig:tetracritical}. If only one condensate were present, the position
of the phase transition would be determined by the point where the respective
$b$ coefficient changes sign. However, when both condensates are present, the
transition lines shift. This is most easily seen from the expression for the
nontrivial solution to the gap equations with both order parameters being
nonzero,
\begin{equation}
\phi_1^2=\frac{-\lambda_{22}b_1+\lambda_{12}b_2}{\lambda_{11}\lambda_{22}-
\lambda_{12}^2}\,,\qquad
\phi_2^2=\frac{\lambda_{12}b_1-\lambda_{11}b_2}{\lambda_{11}\lambda_{22}-
\lambda_{12}^2}\,.
\end{equation}
We can therefore see that the size of the region with both condensates depends
on the sign and magnitude of the offdiagonal coupling $\lambda_{12}$.

The description of the phase transitions based on the GL theory is universal
and model independent as long as it captures the correct degrees of freedom and
symmetries. A nontrivial task in general is to find the mapping of the
$(b_1,b_2)$ plane displayed in Fig.~\ref{Fig:tetracritical} to the physical
observables, in our case the temperature and chemical potential. Even though
this is in principle possible with our PNJL model, in the present work we
performed just a basic compatibility check. Evaluating the GL coefficients for
our parameter set using Eq.~\eqref{GLcoefficients}, one finds that
$\lambda_{\alpha\alpha}\approx2.3\times10^{-3}\Lambda^4$,
$\lambda_{\alpha\Delta}\approx5.7\times10^{-7}\Lambda^2$, and
$\lambda_{\Delta\Delta}\approx9.7\times10^{-6}$. The offdiagonal coupling is
positive which means that the two condensates ``repel'' each other as in the
left panel of Fig.~\ref{Fig:tetracritical}. However, since the GL couplings are
numerically very small, the angles between the critical lines hardly change at
the tetracritical point. The slight deflection of the BEC transition line is
visible in the left panel of Fig.~\ref{Fig:PD2color}. That the same happens to
the deconfinement line is made manifest by the detail of the critical line
shown in the right panel of Fig.~\ref{Fig:PD2color}.


\subsection{Casimir scaling}
\label{Subsec:2cCasimir} The Casimir scaling hypothesis
\cite{Ambjorn:1984mb,DelDebbio:1995gc} states that the color-singlet potential
between a static quark and antiquark at intermediate distance is proportional
to the quadratic Casimir invariant, $C_2(\mathcal R)$, of the representation
$\mathcal R$ of the quarks. This statement is exact at two-loop order in
perturbation theory \cite{Schroder:1998vy} and receives corrections only at
three-loop order \cite{Anzai:2010td}. At the same time, there is compelling
evidence from lattice simulations that it holds to a high accuracy even in the
nonperturbative regime
\cite{Gupta:2007ax,Deldar:1999vi,Bali:2000un,Piccioni:2005un}. It may thus
provide a handle to understand the nonperturbative behavior of QCD-like
theories, and as such should be a necessary ingredient in any model attempting
to mimic QCD (thermo)dynamics \cite{Shevchenko:2000du}.

In the PNJL model, one cannot directly access the confining potential feature
of QCD. However, the scaling of the static potential implies an analogous
property of the expectation values of the Polyakov
loops~\cite{Gupta:2007ax,Meisinger:2001cq}: the quantity $\langle\ell_{\mathcal
R}\rangle^{1/C_2(\mathcal R)}$ should be independent of the representation
$\mathcal R$. Since we have the analytic formula \eqref{2colorpolyakovave} for
the expectation values of all Polyakov loops in two-color QCD, where one has
simply $C_2(j)=j(j+1)$, we can easily check to what extent Casimir scaling is
satisfied by our model.
\FIGURE[t]{%
\includegraphics[scale=0.9]{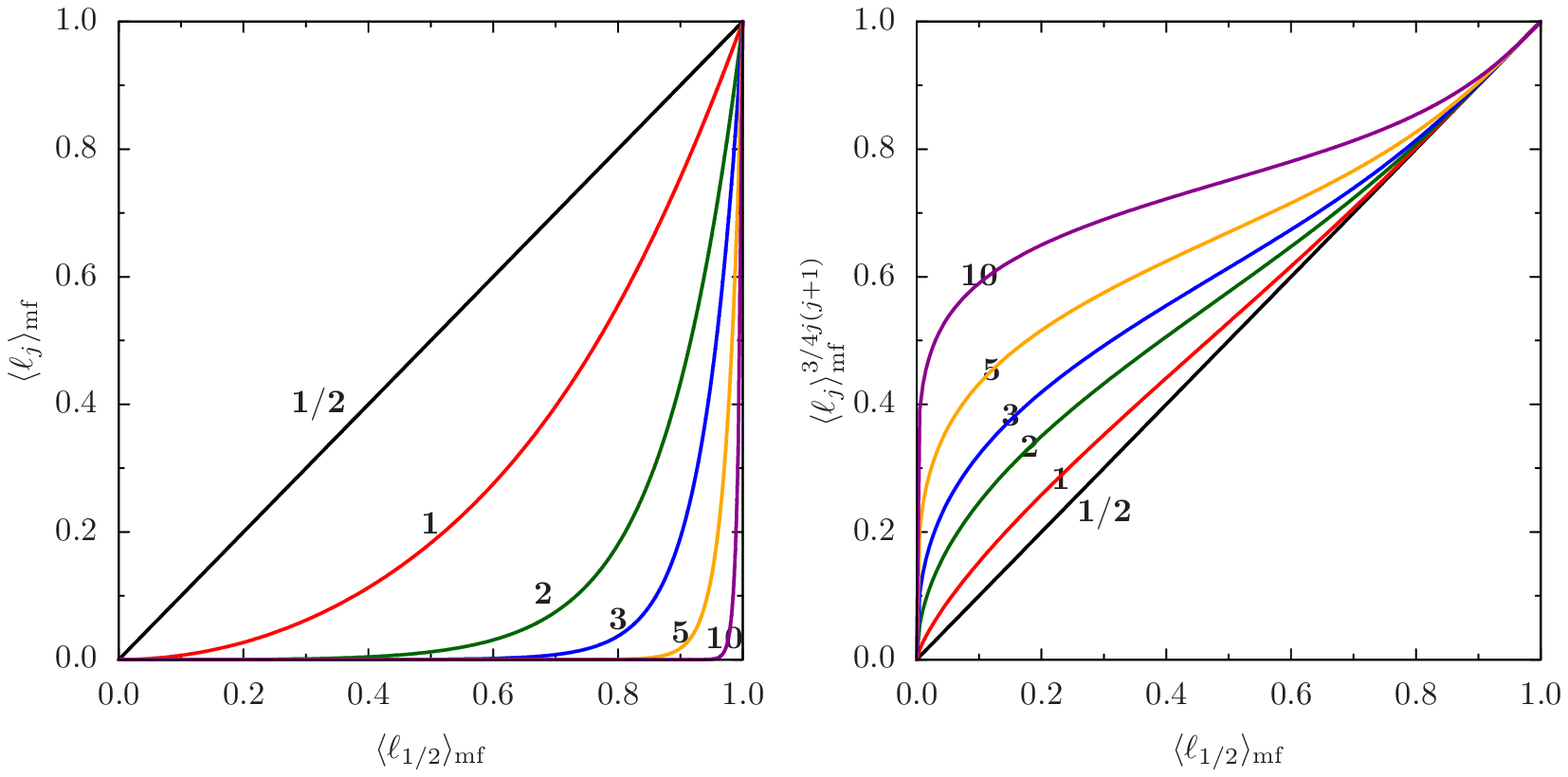}
\caption{Expectation values of the Polyakov loops in various representations as
a function of the fundamental Polyakov loop in the case of two colors. Boldface
numbers indicate the ``spin'' $j$ of the representation. Left panel: unscaled
Polyakov loops. Right panel: Casimir-scaled Polyakov loops. For convenience, we
take the $C_2(\mathrm F)/C_2(\mathcal R)$ power of the expectation values of
the Polyakov loops so that the fundamental loop is left intact.}
\label{Fig:casimir}%
}

Note that the expectation values of all Polyakov loops are expressed in terms
of the mean field $\alpha$, which can in turn be traded for the fundamental
loop. In Fig.~\ref{Fig:casimir} we therefore plot the expectation values of the
Polyakov loops in selected representations against that in the fundamental
representation \cite{Gupta:2007ax,Tsai:2008je}. Comparing the left and right
panels that display the unscaled and scaled Polyakov loops, we can see that the
Casimir scaling is very well reproduced as the value of the fundamental loop
approaches one, which corresponds to high temperatures. It becomes worse at low
temperatures, where the nearest-neighbor interaction model \eqref{eq:sgauge} is
too oversimplified. Lattice data that hint at almost exact scaling even at low
temperatures can be reproduced more satisfactorily once we add more terms
including higher representation Polyakov loops in Eq.~\eqref{eq:sgauge}
\cite{Gupta:2007ax}.

Within our model, we can check even analytically how well Casimir scaling is
satisfied at high temperatures, and hence, at high values of $\alpha$. Carrying
out the Taylor expansion of Eq.~\eqref{2colorpolyakovave} around
$\alpha=+\infty$, one finds
\begin{equation}
\mfave{\ell_j}^{1/j(j+1)}=1-\frac1\alpha+\frac{1}{4\alpha^2}+
\frac{j^2+j-\frac18}{12\alpha^3}+\mathcal O\Bigl(\frac1{\alpha^4}\Bigr)\,.
\end{equation}
We can see that Casimir scaling is only violated at the fourth order of the
expansion.

One important observation regarding our results in Fig.~\ref{Fig:casimir} is
that they are based just on the group average \eqref{groupaverage} and do not
make any reference to the quark sector of the model. Therefore, they apply
equally well to two-color QCD with quarks in \emph{any} representation as well
as to the pure gauge theory. In particular, the same curves hold even for
nonzero chemical potential, which provides us with a unique opportunity to study
Casimir scaling at nonzero baryon density. The quark sector will just affect the
dependence of the mean field $\alpha$ on the temperature and chemical potential,
and therefore the speed at which the curves are traversed as $T$ and $\mu$ vary.


\section{Three colors}
\label{Sec:3c}
For three colors, the group integration is performed with the measure
\begin{equation}
\dd L=\frac{\dd\theta_1\dd\theta_2}{6\pi^2}\bigl[
\sin(\theta_1-\theta_2)-\sin(2\theta_1+\theta_2)+\sin(\theta_1+2\theta_2)
\bigr]^2\,.
\end{equation}
Three-color QCD with fundamental quarks has a charge conjugation invariance,
which is implemented in the PNJL model by a simultaneous change
$\theta_i\to-\theta_i$, $\mu\to-\mu$. Therefore, at any fixed nonzero chemical
potential this charge conjugation invariance is explicitly broken. As a result,
the expectation values $\langle\ell_{\rm F}\rangle$ and $\langle\ell^*_{\rm
F}\rangle$ split. At the same time, the mean-field $\beta$ becomes nonzero
\cite{Abuki:2009gauge}.

On the other hand, the situation in aQCD is different. Thanks to the reality of
the gauge group representation, the nonzero weights appear in pairs with
opposite sign. Consequently, the theory is invariant under \emph{separate}
charge conjugation in the quark and gluon sectors. The charge conjugation
invariance in the gauge sector guarantees that the Polyakov loop in a
given (e.g.~fundamental) representation and its complex conjugate always
have the same expectation value. We may therefore dispense with the mean field
$\beta$, which greatly simplifies the group integration. In the gauge sector one
can still obtain an analytic expression for the thermodynamic potential, albeit
in the form of an infinite series \cite{Fukushima:2003fm}. One defines a
function
\begin{equation}
F(\alpha)=\sum_{m=-\infty}^{+\infty}\det I_{m+i-j}(\alpha)\,,
\label{functionF}
\end{equation}
where the determinant is taken with respect to the indices $i,j$. One then
finds the following expression for the thermodynamic potential,
\begin{equation}
\frac{\Omega^{\rm W}_ga_s^3}{TV}=-6e^{-a/T}\left[\frac{F'(\alpha)}{F(\alpha)}
\right ] ^2
+\alpha\frac{F'(\alpha)}{F(\alpha)}-\log F(\alpha)\,,
\label{3colorOmegagauge}
\end{equation}
and the expectation value of the fundamental Polyakov loop,
\begin{equation}
\mfave{\ell_{\rm F}}=\frac1{N_c}\frac{F'(\alpha)}{F(\alpha)}\,.
\end{equation}
The derivation of this formula is deferred to Appendix
\ref{App:groupintegration} where it will be generalized and used to write
analytic expressions for the expectation values of all Polyakov loops.

The eigenvalues of the Polyakov loop in the adjoint representation are $1$
[$(N_c-1)$-times degenerate] and $e^{\imag(\theta_i-\theta_j)}$ for all pairs
$i\neq j$. The logarithmic term in Eq.~\eqref{Omega1flavor} becomes
\begin{equation}
2\log\biggl\langle(1+x)^{N_c-1}\prod_{i<j}^{N_c}
\bigl[1+2x\cos(\theta_i-\theta_j)+x^2\bigr]\biggr\rangle_{\text{mf}}\,,
\end{equation}
where we abbreviated $x=e^{-E^e_{\bs k}/T}$. Specifically for three colors this
is equal to
\begin{equation}
2\log\Bigl\{(1+x)^2
\bigl[1+2x\omega_1+x^2(3+4\omega_2)
+4x^3(\omega_1+2\omega_3)+x^4(3+4\omega_2)
+2x^5\omega_1+x^6
\bigr]\Bigr\}\,.
\end{equation}
Group integration reduces to evaluation of three averages,
\begin{equation}
\begin{split}
\omega_1=&\mfave{\cos(\theta_1-\theta_2)+\cos(\theta_2-\theta_3)+
\cos(\theta_3-\theta_1)}\,,\\
\omega_2=&\langle\cos(\theta_1-\theta_2)\cos(\theta_3-\theta_1)
+\cos(\theta_2-\theta_3)\cos(\theta_1-\theta_2)
+\cos(\theta_3-\theta_1)\cos(\theta_2-\theta_3)\rangle_{\text{mf}}\,,\\
\omega_3=&\mfave{\cos(\theta_1-\theta_2)\cos(\theta_2-\theta_3)
\cos(\theta_3-\theta_1)}\,.
\end{split}
\label{omegacoeffs}
\end{equation}
These can be performed independently of the value of $x$, so the evaluation of
the quark thermodynamic potential factorizes into a one-dimensional momentum
integral and a two-dimensional group integration. The latter can be performed
either numerically or even analytically in a fashion similar to
Eq.~\eqref{functionF}, as sketched in Appendix \ref{App:groupintegration}.


\subsection{Phase diagram}
\label{Subsec:3c_PD}
\TABLE[t]{%
\setlength{\tabcolsep}{0.7em}
\renewcommand{\arraystretch}{1.5}
\begin{tabular}{ccccc}
\hline\hline
$a\text{ [MeV]}$ & $b^{1/3}\text{ [MeV]}$ & $\Lambda\text{ [MeV]}$
&
$G\text{ [GeV$^{-2}$]}$ & $m_0\text{ [MeV]}$\\
\hline
$542.1$ & $333.2$ & $651$ & $8.51$ & $5.5$\\
\hline\hline
\end{tabular}
\caption{Model parameters for three-color QCD with adjoint quarks.}
\label{Tab:3colorpars}%
} Again, we fix the parameters for the subsequent numerical computations first.
The parameter $a$ is determined by the deconfinement temperature $T_d^0$ in the
pure gauge theory. With the thermodynamic potential \eqref{3colorOmegagauge},
this corresponds to $e^{-a/T_d^0}=0.13427$. Demanding $T_d^0=270\text{ MeV}$,
this yields $a=542.1\text{ MeV}$. The parameter $b$ is in turn obtained from
the physical string tension $\sigma_s=(425\text{ MeV})^2$, as in the two-color
case. In the NJL sector, we use the parameters of the two-flavor model with
fundamental quarks, $\Lambda=651\text{ MeV}$, $G=5.04\text{ GeV}^{-2}$,
$m_0=5.5\text{ MeV}$, fitted to reproduce the pion mass and decay constant and
the chiral condensate in the vacuum (see, for instance,
Ref.~\cite{Ratti:2005phases}). The coupling is rescaled by the factor $27/32$
in accord with Eq.~\eqref{GAGFratio}, and an additional factor of two to
account for the fact that we deal just with one flavor here. The values of all
parameters used in our calculations are summarized in
Tab.~\ref{Tab:3colorpars}.

\FIGURE[t]{%
\includegraphics[scale=0.9]{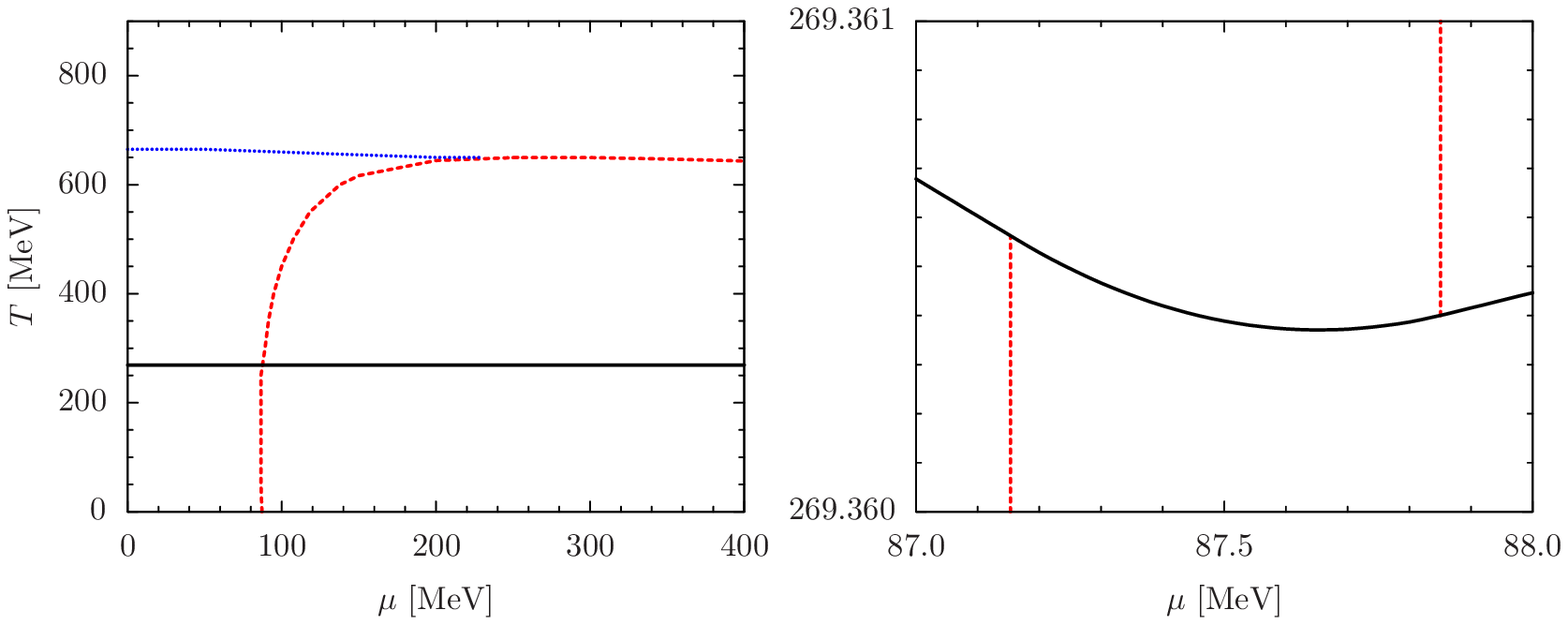}
\caption{Phase diagram of three-color QCD with one flavor of adjoint quarks.
Black solid line: deconfinement transition. Red dashed line: BEC transition.
Blue dotted line: chiral crossover. The right panel zooms in the chemical
potential and temperature scales so that the two tricritical points are
discernible.}
\label{Fig:PD3color}%
}
As a basic cross-check we again evaluated first the deconfinement and chiral
restoration temperatures (in the chiral limit) at zero chemical potential. The
values $T_d=270\text{ MeV}$ and $T_\chi=663\text{ MeV}$ yield the ratio
$T_\chi/T_d=2.46$. This is quite far from the value $\approx8$ measured on the
lattice \cite{Karsch:1998deconfinement,Engels:2005te}. (Note that in
Ref.~\cite{Nishimura:2009me} the lattice value of this ratio was achieved by
tuning the parameters of the model.) However, one should keep in mind that we
made just a rough estimate of the NJL coupling $G$ and cutoff $\Lambda$, on
which the chiral restoration temperature depends very sensitively. In principle,
one could use the lattice value for the ratio $T_\chi/T_d$ as an input in the
model. Nevertheless, one cannot really hope to describe the chiral restoration
in a quantitatively satisfactory manner within our model. The first reason is
that at such high temperatures, the calculation of the thermodynamic potential
is plagued by cutoff artifacts. (We regulate the whole quark contribution to the
thermodynamic potential, including its finite thermal part.) The second reason
is that the PNJL model ceases to be physically appropriate at temperatures about
two to three times $T_d$ \cite{Fukushima:2008phase}, since it does not capture
the correct gauge degrees of freedom, that is, the deconfined transversely
polarized gluons. We are therefore just content with demonstrating that QCD with
adjoint quarks indeed features a large splitting of the deconfinement and chiral
restoration temperatures.

The phase diagram of aQCD determined within our PNJL model is shown in
Fig.~\ref{Fig:PD3color}. While on the large scale it looks the same as the
phase diagram of aQC$_2$D in Fig.~\ref{Fig:PD2color}, there is a marked
difference in the topology as one zooms in the neighborhood of the
``intersection'' of the deconfinement and BEC transition lines. Since the
deconfinement transition is now first order, the BEC critical line is broken,
meeting the deconfinement line at two tricritical points. Thus, there is a
narrow range of chemical potentials in which, as the temperature is increased,
the diquark condensate rather unusually disappears in a first-order phase
transition.


\subsection{Casimir scaling}
\FIGURE[t]{%
\includegraphics[scale=0.9]{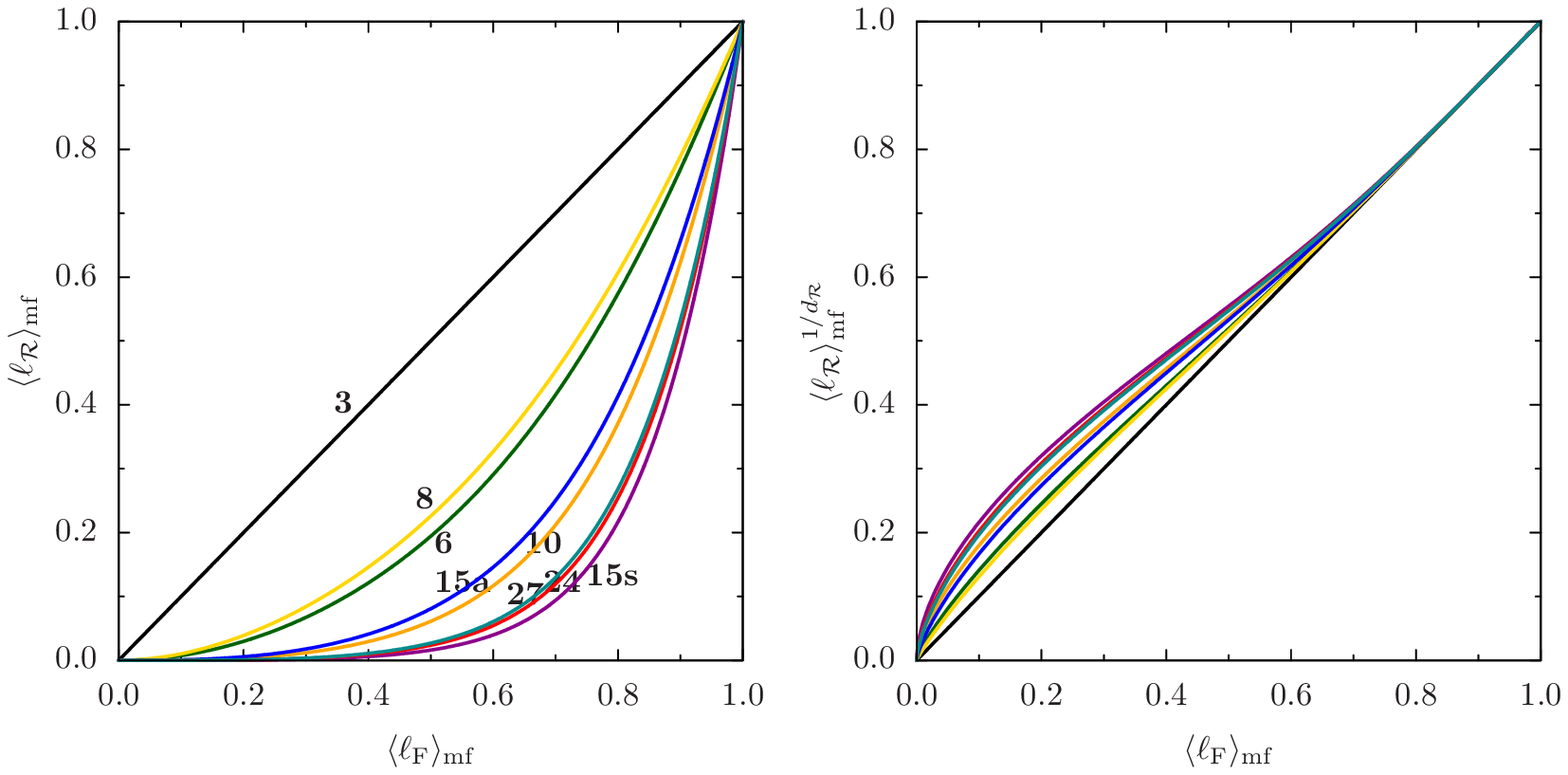}
\caption{Expectation values of the Polyakov loops in various representations as
a function of the fundamental Polyakov loop in the case of three colors.
Boldface numbers indicate the dimension (and possibly the symmetry) of the
representation. Left panel: unscaled Polyakov loops. Right panel:
Casimir-scaled Polyakov loops. For convenience, we take the $C_2(\mathrm
F)/C_2(\mathcal R)\equiv1/d_{\mathcal R}$ power of the expectation values of
the Polyakov loops so that the fundamental loop is left intact. For the sake of
clarity, the labels are not shown in the right panel. The color assignment of
the lines is the same as in the left panel.}
\label{Fig:casimir3c}%
}
Any irreducible representation of $\gr{SU(3)}$ can be uniquely characterized by
a pair of positive integers $(p,q)$ that determine the highest weight of the
representation in the basis of the fundamental weights. The triplet
representation thus corresponds to $(1,0)$ and its complex conjugate to
$(0,1)$. The dimension of a general irreducible representation is
$\dim(p,q)=\frac12(p+1)(q+1)(p+q+2)$ and the value of the quadratic Casimir
invariant (up to a common prefactor) is $C_2(p,q)=\frac13(p^2+pq+q^2)+p+q$
\cite{Cahn:2006ca}. Following Refs.~\cite{Abuki:2009gauge,Bali:2000un}, we have
calculated the expectation values of the Polyakov loops in the lowest few
representations, satisfying $p+q\leq4$. The results are shown in
Fig.~\ref{Fig:casimir3c}.

As before, these results are largely independent of the quark content of the
theory. The only assumption made is that the mean field $\beta$ is zero so that
there is a one-to-one correspondence between the mean field $\alpha$ and the
expectation value of the fundamental Polyakov loop. Thus, the plots in
Fig.~\ref{Fig:casimir3c} apply to three-color QCD modeled by the action
\eqref{eq:sgauge} with quarks in \emph{any} representation at zero chemical
potential. Once the quark representation is (pseudo)real, the same results are
valid even at nonzero chemical potential. As compared to the two-color case
shown in Fig.~\ref{Fig:casimir}, the scaling violation seems to be significantly
smaller for three colors. However, this observation is somewhat misleading since
even the unscaled Polyakov loops show smaller depletion compared to the
fundamental loop in the three-color case.


\section{Conclusions}
\label{Sec:conclusions} In the present paper, we worked out a description of
the thermodynamics of QCD-like theories at nonzero temperature and baryon
chemical potential based on the PNJL model. To mimic the gauge sector we used
a lattice spin model with nearest-neighbor interactions whose parameters are
fixed with the help of the strong-coupling expansion of the full lattice gauge
theory. The quark sector was modeled using the standard NJL model.

We derived a simple mean-field expression for the thermodynamic potential,
which is applicable to QCD-like theories with any number of colors and with
quarks in any representation, as long as this representation is (pseudo)real.
In a sequel to Ref.~\cite{Andersen:2010phase}, we constructed the NJL
Lagrangians for the two classes of QCD-like theories, denoted as type~I and
type~II.

We showed at hand of the example of QCD with adjoint quarks that the Weiss
mean-field approximation to the lattice spin model used here is superior to the
naive mean-field approximation, commonly employed in literature, which leads to
a thermodynamic instability. The Weiss mean-field approximation also allowed us
to derive the expectation value of the Polyakov loop in an arbitrary
representation. The results are given in an implicit form applicable at all
temperatures and chemical potentials, which enables us to study Casimir scaling
in hot and/or dense matter.

As a concrete example, we studied the phase diagram of QCD with adjoint quarks
of two and three colors. We confirmed that in adjoint QCD the critical
temperature for chiral restoration is much higher than that of deconfinement,
both being well-defined phase transitions associated with spontaneous
breaking/restoration of an exact symmetry (the former in the chiral limit). We
checked the model-independent prediction that the phase diagram of aQC$_2$D
features a tetracritical point. On the contrary, in the phase diagram of aQCD
the second-order BEC transition line is interrupted and meets the first-order
deconfinement line at two tricritical points.

It is worth emphasizing that while fine numerical details of our phase diagrams
depend on our guess for the model parameters as well as on the particular way
quarks are implemented, their qualitative features are largely based on
symmetry and thus model-independent. Moreover, our results for Casimir scaling
do not depend on the quark sector, in particular on the choice of the NJL
parameters. They can therefore be understood as a direct test of the lattice
spin model with nearest-neighbor interactions. Once a model for the quark
sector is introduced, they give a prediction for Casimir scaling of
Polyakov-loop expectation values in the whole phase diagram.

In view of the recent lattice data \cite{Hands:2006ve,Hands:2010gd}, the study
of QCD-like theories offers a unique opportunity to gain more insight in the
nature of strongly interacting dense matter. Even though all physical
predictions eventually have to be made within the full gauge theory, we hope to
have demonstrated that the PNJL model provides a versatile tool suitable for
quick calculations and qualitative checks of the robust properties of the
theory.


\acknowledgments The authors are indebted to H.~Abuki and K.~Fukushima for
numerous helpful discussions. Part of the work was carried out during the stay
of T.B.~at the Norwegian University of Science and Technology, Trondheim.
T.B.~would also like to gratefully acknowledge the warm hospitality of the
Department of Energy's Institute for Nuclear Theory at the University of
Washington, where the manuscript was completed. The research of T.Z.~was
supported by the German Academic Exchange Service (DAAD) and by the Helmholtz
Graduate School for Hadron and Ion Research. The research of T.B.~and D.H.R.~was
supported in part by the ExtreMe Matter Institute EMMI in the framework of the
Helmholtz Alliance Program of the Helmholtz Association (HA216/EMMI).


\appendix
\section{Fierz transformation of the current--current interaction}
\label{App:Fierz}
Consider a fermionic field $\psi$ transforming in a representation $\mathcal R$
of the symmetry group. In NJL-like models, one deals with contact four-fermion
interactions of the type $\sum_a(\psib\Gamma^{\mathcal A}_a\psi)^2$, where
$\Gamma^{\mathcal A}_a$ is a set of matrices that project out a particular
irreducible component $\mathcal A$ of the product representation
$\overline{\mathcal R}\otimes\mathcal R$. The Fierz rearrangement of the
four-fermion interaction is equivalent to the group-theoretical identity
\begin{equation}
\sum_a(\Gamma^{\mathcal A}_a)_{ij}(\Gamma^{\mathcal A}_a)_{kl}=\sum_{\mathcal
B}C_{\mathcal{AB}}\sum_b(\Gamma^{\mathcal B}_b)_{il}(\Gamma^{\mathcal
B}_b)_{kj}\,,
\label{fierzaux}
\end{equation}
where the coefficients $C_{\mathcal{AB}}$ depend only on the representations
$\mathcal{A,B}$. In order to fix the effective coupling in the meson channel,
we need not evaluate the Fierz coefficients for all $\mathcal B$. All we need to
know is the coefficient for the one-dimensional representation $\mathcal B=
\mathcal I$, which is always contained in the product $\overline{\mathcal
R}\otimes\mathcal R$.

Setting $\Gamma^{\mathcal I}=\openone$, the coefficient $C_{\mathcal{AI}}$ is
projected out by multiplying Eq.~\eqref{fierzaux} by $\delta_{li}\delta_{jk}$,
which yields
\begin{equation}
C_{\mathcal{AI}}=\frac{\displaystyle\sum_a\tr(\Gamma^{\mathcal
A}_a\Gamma^{\mathcal A}_a)}{(\dim\mathcal R)^2}\,.
\label{fierzaux2}
\end{equation}
In particular for $\mathcal{A=I}$ this leads to
$C_{\mathcal{II}}=1/\dim\mathcal R$. This explains the $1/N_f$ factor in the
effective NJL couplings derived from the current--current interaction
\eqref{current}: both the original interaction as well as the term
$(\psib\psi)^2$ whose coefficient we calculate are in the flavor-singlet
channel. Likewise, the Fierz transformation from the Lorentz-vector channel to
the Lorentz-scalar channel has the Fierz coefficient one.

The color structure of the current--current interaction \eqref{current} is such
that $\mathcal A$ corresponds to the adjoint representation, that is,
$\Gamma^{\mathcal A}_a=T_{a\mathcal R}$ are the generators of the color group
in the representation $\mathcal R$ of the quark fields. The Fierz coefficient
\eqref{fierzaux2} then reduces to $C_{\mathcal{AI}}=C_2(\mathcal
R)/\dim\mathcal R$. Specifically for the $\gr{SU}(N)$ group, once the
generators in the fundamental representation are normalized as $\tr(T_{a\rm F}
T_{b\rm F})=\frac12\delta_{ab}$, one finds $C_2({\rm F})=(N^2-1)/(2N)$ for the
fundamental and $C_2({\rm A})=N$ for the adjoint representation
\cite{Peskin:1995ev}. This concludes the derivation of the effective NJL
couplings $G_{\rm F}$ and $G_{\rm A}$ given below Eq.~\eqref{current}.


\section{Gauge group averaging with continuum quarks}
\label{App:averaging}
In this appendix we justify our prescription \eqref{Omega1flavor} for
adding quarks to the lattice model of the gauge sector. In contrast to
Eq.~\eqref{Omega1flavor}, the authors of Ref.~\cite{Abuki:2009gauge} calculated
the quark thermodynamic potential $\Omega_q$ in the mean-field NJL model with a
constant background gauge field and set $\mfave{\Omega_q}$ as the quark
contribution to the thermodynamic potential.

To start, let us emphasize that any attempt at adding \emph{continuum}
quarks to a lattice gauge model is at best heuristic. For a proper treatment
one would need to discretize the quark action as well, thereby losing the
computational simplicity of the mean-field NJL model. With this in mind, below
we provide a qualitative argument why Eq.~\eqref{Omega1flavor} is a reasonable
approximation.

Imagine adding quarks to the lattice model \eqref{eq:sgauge}; the full action
then formally reads $\Sa=\Sa_g+\psib\Da\psi$, where $\Da$ is the Dirac operator
including the background gauge field the quarks interact with. The full
partition function of the system is obtained as
\begin{equation}
\Za=\int\dd L\,\dd\psi\,\dd\psib\,e^{-\Sa}=\int\dd L\,e^{-\Sa_g}\det\Da\,.
\end{equation}
Using the same trick of introducing the Weiss mean-field action as in
Sec.~\ref{Sec:setup}, this leads to
\begin{equation}
\Za=\Bigl\langle e^{-(\Sa_g-\Sa_{\text{mf}})}\det\Da\Bigr\rangle_{\text{mf}}
\int\dd L\,e^{-\Sa_{\text{mf}}}\,.
\end{equation}
This expression is still exact and includes all correlations between the gauge
and the quark sectors. However, to evaluate it numerically would be very
demanding. We therefore perform a mean-field approximation by setting
\begin{equation}
\Bigl\langle e^{-(\Sa_g-\Sa_{\text{mf}})}\det\Da\Bigr\rangle_{\text{mf}}\approx
e^{-\mfave{\Sa_g-\Sa_{\text{mf}}}}\mfave{\det\Da}\,.
\label{brutalapprox}
\end{equation}

This is equivalent to the Weiss mean-field approximation introduced in
Sec.~\ref{Sec:setup} plus neglecting the correlations between the gauge and
quark sectors.\footnote{The lack of correlations, in particular the feedback
from the dense quark matter into the gauge sector, makes the usual PNJL model
rather trivial in the region of cold dense matter. It would be interesting to
see to what extent these correlations can be taken into account within the
present model.} The full thermodynamic potential is then given by the gauge
part \eqref{weiss} augmented with $-T\log\mfave{\det\Da}$. One can therefore
see that averaging the determinant of the Dirac operator is more natural than
averaging its logarithm. However, Eq.~\eqref{Omega1flavor} commits one more
approximation: it neglects correlations between modes of different momentum and
spin. While the former is naturally incorporated in Eq.~\eqref{Omega1flavor} by
the momentum integral, the latter has to be imposed by hand (by adding the
power $1/2$ to the argument of the logarithm) in presence of a diquark
condensate, since this ties together quarks of opposite spin. Somewhat
ambiguous as this procedure is, it does reproduce the prescription of Abuki and
Fukushima \cite{Abuki:2009gauge} when $\Delta=0$, and, unlike other
prescriptions, it leads to a thermodynamically consistent potential $\Omega_q$
as will now be discussed.

Let us start rather generally by addressing the following question: why have we
used the complicated-looking Weiss mean-field approximation instead of the
simple ``naive'' one?\footnote{We are indebted to Kenji Fukushima for
clarifying this point at the initial stage of the project.} To find the answer
it is useful to understand the relation between the two approximations. Let us
write the Haar measure \eqref{haar} as
\begin{equation}
\dd L=H(\theta)\prod_{i=1}^{N_c-1}\dd\theta_i\,.
\end{equation}
The group integral of a given function $f(\theta)$, weighted by the mean-field
action, can then be expressed as
\begin{equation}
\int\dd L\,f(\theta)e^{-\Sa_{\text{mf}}}=\int\prod_{i=1}^{N_c-1}\dd\theta_i\,
f(\theta)\,e^{-\Sa_{\text{mf}}+\log H(\theta)}\,.
\end{equation}
While in the Weiss mean-field approximation this group integral is evaluated
exactly, the naive mean-field approximation can be obtained by picking the
contribution of the saddle point of the ``action'' $\Sa_{\text{mf}}-\log
H(\theta)$. Indeed, let the saddle point, depending on $\alpha,\beta$, be
$\theta_{\text{mf}}$. Then the above integral is approximated by
$f(\theta_{\text{mf}})e^{-\Sa_{\text{mf}}(\theta_{\text{mf}})+\log
H(\theta_{\text{mf}})}$. The average of any function of the Polyakov loop is
thus simply
\begin{equation}
\mfave{f(\theta)}=f(\theta_{\text{mf}})\,.
\label{naiveaverage}
\end{equation}
Then, in the gauge thermodynamic potential \eqref{weiss}, the Weiss mean fields
$\alpha,\beta$ drop out and the result depends only on $\theta_{\text{mf}}$,
\begin{equation}
\frac{\Omega^{\rm naive}_ga_s^3}{TV}=-2(d-1)N_c^2e^{-a/T}\ell_{\rm
F}(\theta_{\text{mf}})
\ell_{\rm F}^*(\theta_{\text{mf}})-\log H(\theta_{\text{mf}})\,.
\end{equation}
In some particular cases, it can even be expressed solely in terms of the traced
Polyakov loop.

Let us now for simplicity assume that the chemical potential is zero so that
there is no pairing and the Polyakov loop and its complex conjugate give rise to
the same expectation values. The quasiparticle contribution to the quark
thermodynamic potential \eqref{Omega1flavor} with quarks in the representation
$\mathcal R$ of the gauge group then reads
\begin{equation}
-2\threeint{\bs k}\Bigl[\epsilon_{\bs k}\dim\mathcal R
+T\tr\log(\openone+L_{\mathcal R}e^{-\epsilon_{\bs k}/T})\Bigr]\,.
\label{naiveformula}
\end{equation}
The fundamental and adjoint Polyakov loops are related by $\tr L_{\rm
A}=\left|\tr L_{\rm F}\right|^2-1$, hence the same relation holds for their
expectation values in the naive mean-field approximation. This means that at
low temperature when the fundamental Polyakov loop goes to zero, the adjoint
loop should become negative. Disregarding the obvious disagreement of this
conclusion with lattice simulations, it would moreover be a disaster for the
mean-field PNJL model. Indeed, at low temperatures,
\begin{equation}
\tr\log(\openone+L_{\mathcal R}e^{-\epsilon_{\bs k}/T})\approx
e^{-\epsilon_{\bs k}/T}\tr L_{\mathcal R}\,.
\end{equation}
A negative value of the Polyakov loop would thus imply that the quasiquarks
would give a negative contribution to the pressure, leading to a thermodynamic
instability. We conclude that the naive mean-field approximation cannot be
applied to QCD with adjoint quarks.

We will now show that a similar, albeit milder, instability occurs when one
defines the quark contribution to the thermodynamic potential by taking
$\mfave{\Omega_q}$. For the sake of simplicity we focus on aQC$_2$D at low
temperature. The mean field $\alpha$ is then strictly zero (deconfinement is a
sharp phase transition for adjoint quarks) and the average of the quark
thermodynamic potential is easily evaluated using the integrals (14) of
Ref.~\cite{Fukushima:2003fm}. In accord with the general expression
\eqref{Omega1flavor} (with swapped logarithm and averaging operations), one
finds
\begin{equation}
2\left\langle
\log\bigl[(1+x)(1+2x\cos2\theta+x^2)\bigr]
\right\rangle_{\text{mf}}
=2[\log(1+x)-x]\,,
\label{wrongeq}
\end{equation}
where $x=e^{-E^e_{\bs k}/T}$. Even though the leading term, linear in $x$ and
proportional to $\mfave{\tr L_{\rm A}}$, now vanishes, the total quasiquark pressure is
still negative. This negative contribution is numerically small, yet it makes
the thermodynamics in principle ill-defined.

It is easy to see that this problem does not arise when the group average is
taken inside the logarithm as in Eq.~\eqref{Omega1flavor}. Then at low
temperature when $\alpha=0$, one gets instead of Eq.~\eqref{wrongeq}
\begin{equation}
2\log\bigl\langle
(1+x)(1+2x\cos2\theta+x^2)
\bigr\rangle_{\text{mf}}
=2\log\bigl[(1+x)(1-x+x^2)\bigr]=2\log(1+x^3)\,.
\label{correcteq}
\end{equation}
The pressure is now strictly positive and even looks like a pressure of
noninteracting fermionic quasiparticles with energy $3E^e_{\bs k}$.

One comment is appropriate regarding the last claim. In the PNJL model for
physical, three-color QCD with fundamental quarks, one observes the same
behavior at low temperature. More precisely, the mean field $\alpha$ is never
strictly zero at any nonzero temperature, so the quark contribution to the
pressure is proportional to $\log(1+3x\ell_{\rm F}+3x^2\ell^*_{\rm
F}+x^3)$. At low temperature when the Polyakov loop is suppressed this reduces
to $\log(1+x^3)$, which is usually interpreted as a manifestation of the fact
that one needs three quarks to create a color-singlet state. This observation
suggests that the PNJL model is a natural framework for a description of the
quarkyonic phase in cold dense quark
matter~\cite{McLerran:2007qj,Schaefer:2007pw,Abuki:2008nm}. However, as
Eq.~\eqref{correcteq} clearly shows, this is somewhat misleading: the same
low-temperature behavior of the pressure arises in \emph{two-color} QCD with
adjoint quarks, so it does not directly reflect the number of quarks needed to
construct a color singlet.

A second attempt at interpreting $\log(1+x^3)$ might be that both examples of
three-color fundamental and two-color adjoint quarks are governed by the
dimension of the representation. However, in two-color QCD it is easy to
calculate the same quantity with quarks in higher representations, showing that
there is no simple general relation between the representation and the form of
the low-temperature pressure. For instance, in aQCD below the deconfinement
temperature, the coefficients $\omega_{1,2,3}$ take on the values $\omega_1=-1$,
$\omega_2=0$, $\omega_3=1/8$. Consequently, the quark pressure is proportional
to $2\log(1+x^3+x^5+x^8)$.


\section{Group integration for $\gr{SU}(N)$}
\label{App:groupintegration}
In this appendix we show that some of the group integrals can be performed for
arbitrary $N$~\cite{Creutz:1978ub,Kogut:1981ez}. (For the sake of legibility,
we abbreviate $N_c$ as $N$.) Let us define the generating function
\begin{equation}
\Ga(z,\bar z)=\biggl\langle
\prod_{i=1}^{N}e^{ze^{\imag\theta_i}}e^{\bar ze^{-\imag\theta_i}}
\biggr\rangle_{\text{mf}}\,.
\label{generatingfunction}
\end{equation}
In order to calculate it, we write the mean-field action
\eqref{generalMFaction} for one lattice site as
\begin{equation}
\Sa_{\text{mf}}=-\sum_{i=1}^{N}(\alpha\cos\theta_i+\imag\beta\sin\theta_i)\,.
\end{equation}
Furthermore, we use the fact that the Haar measure \eqref{haar} may be
written as a square of a Vandermonde determinant,
\begin{equation}
\dd L=\prod_{i=1}^N\dd\theta_i\,\delta(\theta_1+\dotsb+\theta_N)
\varepsilon_{i_1\dotsb i_N}\varepsilon_{j_1\dotsb j_N}
e^{\imag\theta_1(i_1-j_1)}\dotsb e^{\imag\theta_N(i_N-j_N)}\,.
\label{vandermonde}
\end{equation}
The last trick is to express the (periodic) $\delta$-function in terms of its
Fourier series,
\begin{equation}
\delta(\theta_1+\dotsb+\theta_N)=\frac1{2\pi}\sum_{m=-\infty}^{+\infty}
e^{\imag m(\theta_1+\dotsb+\theta_N)}\,.
\label{periodicdelta}
\end{equation}
The integration over the angles $\theta_i$ now completely factorizes in terms of
a single master integral,
\begin{equation}
\Ta_n(u,v)=\frac1{2\pi}\int_0^{2\pi}\dd\theta\,e^{\imag
n\theta}e^{u\cos\theta+\imag v\sin\theta}\,,
\label{masterintegral}
\end{equation}
For real $u$ and pure imaginary $v$, $v=\imag w$, which is the case if
$\beta=0$, the master integral can again be expressed with the help of the
modified Bessel function,
\begin{equation}
\Ta_n(u,\imag w)=\biggl(\frac{u-\imag w}{\sqrt{u^2+w^2}}\biggr)^n
I_n\bigl(\sqrt{u^2+w^2}\bigr)\,.
\end{equation}
The final formula for the generating function \eqref{generatingfunction}
reads
\begin{equation}
\Ga(z,\bar z)=\frac{\displaystyle\sum_{m=-\infty}^{+\infty}\det\Ta_{m+i-j}
(\alpha+z+\bar z,\beta+z-\bar z)}
{\displaystyle\sum_{m=-\infty}^{+\infty}\det\Ta_{m+i-j}(\alpha,\beta)}\,.
\label{masterformula}
\end{equation}

Looking back at Eq.~\eqref{generatingfunction} one sees that expanding the
exponentials, the Taylor coefficient of the $z^m\bar z^n$ term resums all
eigenvalues of the Polyakov loop in the ${\rm F}^m\otimes{\rm\bar F}^n$
representation, $\rm F$ being the fundamental one. That is, one has
\begin{equation}
\mfave{\tr L_{{\rm F}^m\otimes{\rm\bar F}^n}}=
\left.\frac{\partial^{m+n}}{\partial z^m\partial\bar z^n}\Ga(z,\bar z)
\right|_{\substack{z=0\\ \bar z=0}}\,.
\end{equation}
The expectation values of Polyakov loops in all \emph{irreducible}
representations can be obtained from this formula by simply observing that the
(traced) Polyakov loop in a direct sum of two representations is equal to the
sum of the loops in these representations.

Let us remark here that the thermodynamic potential of the three-color pure
gauge theory \eqref{3colorOmegagauge} can be derived using the same argument,
and the group integrals involved are special cases of those considered above.
Indeed, the function $F(\alpha)$ \eqref{functionF} equals the denominator in
Eq.~\eqref{masterformula} at $\beta=0$ up to a trivial numerical prefactor.
Changing this prefactor just shifts the thermodynamic potential by a constant,
and noting that $F(0)=1$, it can be fixed by demanding that $\Omega_g=0$ for
$\alpha=0$.

A more compact formula can again be obtained for the special case of
two colors. Then, we can set $\beta=0$ and $\bar z=0$. Also,
$\prod_{i=1}^2e^{ze^{\imag\theta}}=e^{2z\cos\theta}$. The one-dimensional group
integration can be performed directly and one finds
\begin{equation}
\Ga(z)=\frac{I_0(2\alpha+2z)-I_2(2\alpha+2z)}{I_0(2\alpha)-I_2(2\alpha)}=
\frac{\alpha}{\alpha+z}\frac{I_1(2\alpha+2z)}{I_1(2\alpha)}\,.
\end{equation}
While the latter expression is more compact, the former is more convenient for
taking the derivatives in order to extract the expectation values of the
Polyakov loops.

Finally, let us show that even the averages \eqref{omegacoeffs} can be
expressed analytically in terms of a series of modified Bessel functions
\cite{Damgaard:1987wh}, and thus speed up the numerical evaluation of the
thermodynamic potential. Using trigonometric identities, these averages can be
written as a linear combination of terms of the type
\begin{equation}
K_{abc}(\alpha)=\mfave{e^{\imag(a\theta_1+b\theta_2+c\theta_3)}}\,,
\end{equation}
where $a,b,c$ are integers. Using the same trick of rewriting the Haar measure
as a Vandermonde determinant and introducing the periodic $\delta$-function as
in Eqs.~\eqref{vandermonde} and \eqref{periodicdelta}, this becomes
\begin{equation}
K_{abc}(\alpha)=\frac1{6F(\alpha)}\sum_{m=-\infty}^{+\infty}\sum_{i,j,k=1}^3
\varepsilon_{ijk}\left|
\begin{array}{ccc}
I_{m+i-1+a}(\alpha) & I_{m+i-2+a}(\alpha) & I_{m+i-3+a}(\alpha)\\
I_{m+j-1+b}(\alpha) & I_{m+j-2+b}(\alpha) & I_{m+j-3+b}(\alpha)\\
I_{m+k-1+c}(\alpha) & I_{m+k-2+c}(\alpha) & I_{m+k-3+c}(\alpha)
\end{array}
\right|.
\end{equation}


\bibliography{refs_new}

\providecommand{\href}[2]{#2}\begingroup\raggedright\begin{thebibliography}{10}

\bibitem{Alford:1998imaginary}
M.~G. Alford, A.~Kapustin, and F.~Wilczek, {\it {Imaginary chemical potential
  and finite fermion density on the lattice}},  {\em Phys. Rev.} {\bf D59}
  (1999) 054502, [\href{http://arxiv.org/abs/hep-lat/9807039}{{\tt
  hep-lat/9807039}}].

\bibitem{Son:2000qcd}
D.~T. Son and M.~A. Stephanov, {\it {QCD at Finite Isospin Density}},  {\em
  Phys. Rev. Lett.} {\bf 86} (2001) 592--595,
  [\href{http://arxiv.org/abs/hep-ph/0005225}{{\tt hep-ph/0005225}}].

\bibitem{Kogut:1999on}
J.~B. Kogut, M.~A. Stephanov, and D.~Toublan, {\it {On two-color QCD with
  baryon chemical potential}},  {\em Phys. Lett.} {\bf B464} (1999) 183--191,
  [\href{http://arxiv.org/abs/hep-ph/9906346}{{\tt hep-ph/9906346}}].

\bibitem{Kogut:2000qcdlike}
J.~B. Kogut, M.~A. Stephanov, D.~Toublan, J.~J.~M. Verbaarschot, and
  A.~Zhitnitsky, {\it {QCD-like theories at finite baryon density}},  {\em
  Nucl. Phys.} {\bf B582} (2000) 477--513,
  [\href{http://arxiv.org/abs/hep-ph/0001171}{{\tt hep-ph/0001171}}].

\bibitem{Peskin:1980gc}
M.~E. Peskin, {\it {The alignment of the vacuum in theories of technicolor}},
  {\em Nucl. Phys.} {\bf B175} (1980) 197--233.

\bibitem{Bijnens:2009qm}
J.~Bijnens and J.~Lu, {\it {Technicolor and other QCD-like theories at
  next-to-next-to-leading order}},  {\em J. High Energy Phys.} {\bf 11} (2009)
  116, [\href{http://arxiv.org/abs/0910.5424}{{\tt arXiv:0910.5424}}].

\bibitem{Kondratyuk:1991hf}
L.~A. Kondratyuk, M.~M. Giannini, and M.~I. Krivoruchenko, {\it {The
  $\gr{SU}(2)$ colour superconductivity}},  {\em Phys. Lett.} {\bf B269} (1991)
  139--143.

\bibitem{Ratti:2004thermodynamics}
C.~Ratti and W.~Weise, {\it {Thermodynamics of two-color QCD and the Nambu
  Jona-Lasinio model}},  {\em Phys. Rev.} {\bf D70} (2004) 054013,
  [\href{http://arxiv.org/abs/hep-ph/0406159}{{\tt hep-ph/0406159}}].

\bibitem{Sun:2007fc}
G.-f. Sun, L.~He, and P.~Zhuang, {\it {BEC-BCS crossover in the
  Nambu--Jona-Lasinio model of QCD}},  {\em Phys. Rev.} {\bf D75} (2007)
  096004, [\href{http://arxiv.org/abs/hep-ph/0703159}{{\tt hep-ph/0703159}}].

\bibitem{Brauner:2009twocolor}
T.~Brauner, K.~Fukushima, and Y.~Hidaka, {\it {Two-color quark matter:
  $\gr{U}(1)_A$ restoration, superfluidity, and quarkyonic phase}},  {\em Phys.
  Rev.} {\bf D80} (2009) 074035, [\href{http://arxiv.org/abs/0907.4905}{{\tt
  arXiv:0907.4905}}].

\bibitem{Andersen:2010phase}
J.~O. Andersen and T.~Brauner, {\it {Phase diagram of two-color quark matter at
  nonzero baryon and isospin density}},  {\em Phys. Rev.} {\bf D81} (2010)
  096004, [\href{http://arxiv.org/abs/1001.5168}{{\tt arXiv:1001.5168}}].

\bibitem{Nishimura:2009me}
H.~Nishimura and M.~C. Ogilvie, {\it {Polyakov--Nambu--Jona-Lasinio model for
  adjoint fermions with periodic boundary conditions}},  {\em Phys. Rev.} {\bf
  D81} (2010) 014018, [\href{http://arxiv.org/abs/0911.2696}{{\tt
  arXiv:0911.2696}}].

\bibitem{Mocsy:2003qw}
{\'A}.~M{\'o}csy, F.~Sannino, and K.~Tuominen, {\it {Confinement versus Chiral
  Symmetry}},  {\em Phys. Rev. Lett.} {\bf 92} (2004) 182302,
  [\href{http://arxiv.org/abs/hep-ph/0308135}{{\tt hep-ph/0308135}}].

\bibitem{Sannino:2004tetracritical}
F.~Sannino and K.~Tuominen, {\it {Tetracritical behavior in strongly
  interacting theories}},  {\em Phys. Rev.} {\bf D70} (2004) 034019,
  [\href{http://arxiv.org/abs/hep-ph/0403175}{{\tt hep-ph/0403175}}].

\bibitem{Gocksch:1984xc}
A.~Gocksch and M.~Ogilvie, {\it {An effective strong coupling lattice model for
  finite temperature QCD}},  {\em Phys. Lett.} {\bf B141} (1984) 407--410.

\bibitem{Dumitru:2003deconfining}
A.~Dumitru, Y.~Hatta, J.~Lenaghan, K.~Orginos, and R.~D. Pisarski, {\it
  {Deconfining phase transition as a matrix model of renormalized Polyakov
  loops}},  {\em Phys. Rev.} {\bf D70} (2004) 034511,
  [\href{http://arxiv.org/abs/hep-th/0311223}{{\tt hep-th/0311223}}].

\bibitem{Fukushima:2007a}
K.~Fukushima and Y.~Hidaka, {\it {Model study of the sign problem in the
  mean-field approximation}},  {\em Phys. Rev.} {\bf D75} (2007) 036002,
  [\href{http://arxiv.org/abs/hep-ph/0610323}{{\tt hep-ph/0610323}}].

\bibitem{Fukushima:2003fm}
K.~Fukushima, {\it {Relation between the Polyakov loop and the chiral order
  parameter at strong coupling}},  {\em Phys. Rev.} {\bf D68} (2003) 045004,
  [\href{http://arxiv.org/abs/hep-ph/0303225}{{\tt hep-ph/0303225}}].

\bibitem{Gupta:2007ax}
S.~Gupta, K.~{H\"ubner}, and O.~Kaczmarek, {\it {Renormalized Polyakov loops in
  many representations}},  {\em Phys. Rev.} {\bf D77} (2008) 034503,
  [\href{http://arxiv.org/abs/0711.2251}{{\tt arXiv:0711.2251}}].

\bibitem{Abuki:2009gauge}
H.~Abuki and K.~Fukushima, {\it {Gauge dynamics in the PNJL model: Color
  neutrality and Casimir scaling}},  {\em Phys. Lett.} {\bf B676} (2009)
  57--62, [\href{http://arxiv.org/abs/0901.4821}{{\tt arXiv:0901.4821}}].

\bibitem{Fukushima:2004chiral}
K.~Fukushima, {\it {Chiral effective model with the Polyakov loop}},  {\em
  Phys. Lett.} {\bf B591} (2004) 277--284,
  [\href{http://arxiv.org/abs/hep-ph/0310121}{{\tt hep-ph/0310121}}].

\bibitem{Megias:2004hj}
E.~Meg{\'\i}as, E.~Ruiz~Arriola, and L.~L. Salcedo, {\it {Polyakov loop in
  chiral quark models at finite temperature}},  {\em Phys. Rev.} {\bf D74}
  (2006) 065005, [\href{http://arxiv.org/abs/hep-ph/0412308}{{\tt
  hep-ph/0412308}}].

\bibitem{Ratti:2005phases}
C.~Ratti, M.~A. Thaler, and W.~Weise, {\it {Phases of QCD: Lattice
  thermodynamics and a field theoretical model}},  {\em Phys. Rev.} {\bf D73}
  (2006) 014019, [\href{http://arxiv.org/abs/hep-ph/0506234}{{\tt
  hep-ph/0506234}}].

\bibitem{Roessner:2006polyakov}
S.~{R\"o\ss ner}, C.~Ratti, and W.~Weise, {\it {Polyakov loop, diquarks, and
  the two-flavor phase diagram}},  {\em Phys. Rev.} {\bf D75} (2007) 034007,
  [\href{http://arxiv.org/abs/hep-ph/0609281}{{\tt hep-ph/0609281}}].

\bibitem{Fukushima:2008phase}
K.~Fukushima, {\it {Phase diagrams in the three-flavor Nambu--Jona-Lasinio
  model with the Polyakov loop}},  {\em Phys. Rev.} {\bf D77} (2008) 114028,
  [\href{http://arxiv.org/abs/0803.3318}{{\tt arXiv:0803.3318}}].

\bibitem{Kogut:1985xa}
J.~B. Kogut, J.~Polonyi, H.~W. Wyld, and D.~K. Sinclair, {\it {Hierarchical
  Mass Scales in Lattice Gauge Theories with Dynamical, Light Fermions}},  {\em
  Phys. Rev. Lett.} {\bf 54} (1985) 1980--1982.

\bibitem{Karsch:1998deconfinement}
F.~Karsch and M.~{L\"utgemeier}, {\it {Deconfinement and chiral symmetry
  restoration in an $\gr{SU}(3)$ gauge theory with adjoint fermions}},  {\em
  Nucl. Phys.} {\bf B550} (1999) 449--464,
  [\href{http://arxiv.org/abs/hep-lat/9812023}{{\tt hep-lat/9812023}}].

\bibitem{Engels:2005te}
J.~Engels, S.~Holtmann, and T.~Schulze, {\it {Scaling and Goldstone effects in
  a QCD with two flavours of adjoint quarks}},  {\em Nucl. Phys.} {\bf B724}
  (2005) 357--379, [\href{http://arxiv.org/abs/hep-lat/0505008}{{\tt
  hep-lat/0505008}}].

\bibitem{Unsal:2007vu}
M.~{\"U}nsal, {\it {Abelian Duality, Confinement, and Chiral-Symmetry Breaking
  in a $\gr{SU}(2)$ QCD-like Theory}},  {\em Phys. Rev. Lett.} {\bf 100} (2008)
  032005, [\href{http://arxiv.org/abs/0708.1772}{{\tt arXiv:0708.1772}}].

\bibitem{Lohwater:1982lo}
A.~Lohwater, {\em {\href{http://www.mediafire.com/?1mw1tkgozzu} {Introduction
  to Inequalities}}}.
\newblock Online e-book in PDF format, 1982.

\bibitem{Georgi:1982jb}
H.~Georgi, {\em {Lie Algebras in Particle Physics}}.
\newblock Frontiers in Physics. Perseus Books, Reading, Massachusetts,
  second~ed., 1999.

\bibitem{Hands:2000numerical}
S.~Hands, I.~Montvay, S.~Morrison, M.~Oevers, L.~Scorzato, and J.-I. Skullerud,
  {\it {Numerical study of dense adjoint matter in two color QCD}},  {\em Eur.
  Phys. J.} {\bf C17} (2000) 285--302,
  [\href{http://arxiv.org/abs/hep-lat/0006018}{{\tt hep-lat/0006018}}].

\bibitem{Fukushima:2007bj}
K.~Fukushima and K.~Iida, {\it {Larkin--Ovchinnikov--Fulde--Ferrell state in
  two-color quark matter}},  {\em Phys. Rev.} {\bf D76} (2007) 054004,
  [\href{http://arxiv.org/abs/0705.0792}{{\tt arXiv:0705.0792}}].

\bibitem{Splittorff:2000mm}
K.~Splittorff, D.~T. Son, and M.~A. Stephanov, {\it {QCD-like theories at
  finite baryon and isospin density}},  {\em Phys. Rev.} {\bf D64} (2001)
  016003, [\href{http://arxiv.org/abs/hep-ph/0012274}{{\tt hep-ph/0012274}}].

\bibitem{Vafa:1984xg}
C.~Vafa and E.~Witten, {\it {Parity Conservation in Quantum Chromodynamics}},
  {\em Phys. Rev. Lett.} {\bf 53} (1984) 535--536.

\bibitem{Buballa:2003qv}
M.~Buballa, {\it {NJL model analysis of quark matter at large density}},  {\em
  Phys. Rep.} {\bf 407} (2005) 205--376,
  [\href{http://arxiv.org/abs/hep-ph/0402234}{{\tt hep-ph/0402234}}].

\bibitem{Ambjorn:1984mb}
J.~Ambj{\o}rn, P.~Olesen, and C.~Peterson, {\it {Stochastic confinement and
  dimensional reduction: (I). Four-dimensional $\gr{SU}(2)$ lattice gauge
  theory}},  {\em Nucl. Phys.} {\bf B240} (1984) 189--212.

\bibitem{DelDebbio:1995gc}
L.~Del~Debbio, M.~Faber, J.~Greensite, and {\v{S}}.~{Olejn\'{\i}k}, {\it
  {Casimir scaling versus Abelian dominance in QCD string formation}},  {\em
  Phys. Rev.} {\bf D53} (1996) 5891--5897,
  [\href{http://arxiv.org/abs/hep-lat/9510028}{{\tt hep-lat/9510028}}].

\bibitem{Schroder:1998vy}
Y.~{Schr\"oder}, {\it {The static potential in QCD to two loops}},  {\em Phys.
  Lett.} {\bf B447} (1999) 321--326,
  [\href{http://arxiv.org/abs/hep-ph/9812205}{{\tt hep-ph/9812205}}].

\bibitem{Anzai:2010td}
C.~Anzai, Y.~Kiyo, and Y.~Sumino, {\it {Violation of Casimir Scaling for Static
  QCD Potential at Three-loop Order}},
  \href{http://arxiv.org/abs/1004.1562}{{\tt arXiv:1004.1562}}.

\bibitem{Deldar:1999vi}
S.~Deldar, {\it {Static $\gr{SU}(3)$ potentials for sources in various
  representations}},  {\em Phys. Rev.} {\bf D62} (2000) 034509,
  [\href{http://arxiv.org/abs/hep-lat/9911008}{{\tt hep-lat/9911008}}].

\bibitem{Bali:2000un}
G.~S. Bali, {\it {Casimir scaling of $\gr{SU}(3)$ static potentials}},  {\em
  Phys. Rev.} {\bf D62} (2000) 114503,
  [\href{http://arxiv.org/abs/hep-lat/0006022}{{\tt hep-lat/0006022}}].

\bibitem{Piccioni:2005un}
C.~Piccioni, {\it {Casimir scaling in $\gr{SU}(2)$ lattice gauge theory}},
  {\em Phys. Rev.} {\bf D73} (2006) 114509,
  [\href{http://arxiv.org/abs/hep-lat/0503021}{{\tt hep-lat/0503021}}].

\bibitem{Shevchenko:2000du}
V.~I. Shevchenko and Y.~A. Simonov, {\it {Casimir Scaling as a Test of QCD
  Vacuum Models}},  {\em Phys. Rev. Lett.} {\bf 85} (2000) 1811--1814,
  [\href{http://arxiv.org/abs/hep-ph/0001299}{{\tt hep-ph/0001299}}].

\bibitem{Meisinger:2001cq}
P.~N. Meisinger, T.~R. Miller, and M.~C. Ogilvie, {\it {Phenomenological
  equations of state for the quark-gluon plasma}},  {\em Phys. Rev.} {\bf D65}
  (2002) 034009, [\href{http://arxiv.org/abs/hep-ph/0108009}{{\tt
  hep-ph/0108009}}].

\bibitem{Tsai:2008je}
H.-M. Tsai and B.~{M\"uller}, {\it {Phenomenology of the three-flavor PNJL
  model and thermal strange quark production}},  {\em J. Phys.} {\bf G36}
  (2009) 075101, [\href{http://arxiv.org/abs/0811.2216}{{\tt
  arXiv:0811.2216}}].

\bibitem{Cahn:2006ca}
R.~N. Cahn, {\em {Semi-Simple Lie Algebras and Their Representations}}.
\newblock Dover Publications, New York, 2006.

\bibitem{Hands:2006ve}
S.~Hands, S.~Kim, and J.-I. Skullerud, {\it {Deconfinement in dense two-color
  QCD}},  {\em Eur. Phys. J.} {\bf C48} (2006) 193--206,
  [\href{http://arxiv.org/abs/hep-lat/0604004}{{\tt hep-lat/0604004}}].

\bibitem{Hands:2010gd}
S.~Hands, S.~Kim, and J.-I. Skullerud, {\it {Quarkyonic phase in dense two
  color matter}},  {\em Phys. Rev.} {\bf D81} (2010) 091502,
  [\href{http://arxiv.org/abs/1001.1682}{{\tt arXiv:1001.1682}}].

\bibitem{Peskin:1995ev}
M.~E. Peskin and D.~V. Schroeder, {\em {An Introduction to Quantum Field
  Theory}}.
\newblock Addison-Wesley, Reading, Massachusetts, 1995.

\bibitem{McLerran:2007qj}
L.~McLerran and R.~D. Pisarski, {\it {Phases of dense quarks at large $N_c$}},
  {\em Nucl. Phys.} {\bf A796} (2007) 83--100,
  [\href{http://arxiv.org/abs/0706.2191}{{\tt arXiv:0706.2191}}].

\bibitem{Schaefer:2007pw}
B.-J. Schaefer, J.~M. Pawlowski, and J.~Wambach, {\it {Phase structure of the
  Polyakov-quark-meson model}},  {\em Phys. Rev.} {\bf D76} (2007) 074023,
  [\href{http://arxiv.org/abs/0704.3234}{{\tt arXiv:0704.3234}}].

\bibitem{Abuki:2008nm}
H.~Abuki, R.~Anglani, R.~Gatto, G.~Nardulli, and M.~Ruggieri, {\it {Chiral
  crossover, deconfinement, and quarkyonic matter within a Nambu--Jona-Lasinio
  model with the Polyakov loop}},  {\em Phys. Rev.} {\bf D78} (2008) 034034,
  [\href{http://arxiv.org/abs/0805.1509}{{\tt arXiv:0805.1509}}].

\bibitem{Creutz:1978ub}
M.~Creutz, {\it {On invariant integration over $\gr{SU}(N)$}},  {\em J. Math.
  Phys.} {\bf 19} (1978) 2043--2046.

\bibitem{Kogut:1981ez}
J.~B. Kogut, M.~Snow, and M.~Stone, {\it {Mean field and Monte Carlo studies of
  $\gr{SU}(N)$ chiral models in three dimensions}},  {\em Nucl. Phys.} {\bf
  B200} (1982) 211--231.

\bibitem{Damgaard:1987wh}
P.~H. Damgaard, {\it {The free energy of higher representation sources in
  lattice gauge theories}},  {\em Phys. Lett.} {\bf B194} (1987) 107--113.

\end{thebibliography}\endgroup
\bibliographystyle{JHEP}

\end{document}